\documentclass[apj]{emulateapj-rtx4}

\usepackage{amsmath}
\usepackage{amssymb}
\usepackage{amsfonts}
\usepackage{amstext}
\usepackage{graphicx}
\usepackage{fancybox}
\usepackage[usenames,dvipsnames]{color}
\usepackage{pstricks}
\usepackage{bm}
\usepackage{soul,color}
\usepackage{latexsym}
\usepackage{pifont}
\usepackage{shorttoc}
\usepackage{subfigure}
\usepackage{textcomp} 
\usepackage{float}
\usepackage{enumitem}

\usepackage{bm}
\usepackage{lipsum}
\usepackage{threeparttablex}
\usepackage{hyperref}
\hypersetup{urlcolor=cyan}

\def\aap{A\&A}

\def\apjs{ApJS}
\def\apjl{ApJL}

\newcommand{\lya}{Ly$\alpha$\ }

\newcommand{\no}[1]{}
\renewcommand{\exp}[1]{\mathrm{exp}\left(#1\right)}

\newcommand{\myemail}{lluis.mas-ribas@jpl.nasa.gov}
\def\lsim{~\rlap{$<$}{\lower 1.0ex\hbox{$\sim$}}}

\def\gsim{~\rlap{$>$}{\lower 1.0ex\hbox{$\sim$}}}

\shorttitle{Ubiquitous Radiative Acceleration in Quasar Outflows}
\shortauthors{Mas-Ribas \& Mauland}

\begin{document}

\title{The Ubiquitous Imprint of Radiative Acceleration in the  
Mean \\  Absorption Spectrum of Quasar Outflows}

 \author{Llu\'is Mas-Ribas\altaffilmark{1,2,3}} 
\author{Renate Mauland\altaffilmark{3}} 
\altaffiltext{1}{Jet Propulsion Laboratory, California Institute of Technology, 4800 Oak Grove Drive, Pasadena, CA 91109, U.S.A.}
\altaffiltext{2}{California Institute of Technology, 1200 E. California Blvd, Pasadena, CA 91125, U.S.A.}
\altaffiltext{3}{Institute of Theoretical Astrophysics, University of Oslo,
Postboks 1029, 0315 Oslo, Norway.\\
\url{\myemail}\\
\\
\\
\copyright 2019. All rights reserved.}


\begin{abstract}  

Observational evidence revealing the main mechanisms that accelerate quasar outflows 
has proven difficult to obtain, due to the complexity of the absorption features that this gas 
produces in the spectra of the emission sources. We build 36 composite outflow spectra,  
covering a large range of outflow and quasar parameters, by stacking broad ($>450\,{\rm 
km\,s^{-1}}$) absorption line systems in the spectra of SDSS-III/BOSS DR12 quasars. 
The two lines of the atomic doublet of C{\sc iv}, with a separation of $\approx 497\,{\rm km\,s^{-1}}$,  
as well as those of other species appear well resolved in most of our composites. This agrees 
with broad outflow troughs consisting in the superposition of narrow absorbers. We also 
report on the ubiquitous detection of the radiative-acceleration signature known as line 
locking in all our composite outflow spectra, including one spectrum 
strictly built from broad absorption line (BAL) systems. This is the first line-locking detection 
in BAL composite spectra. Line locking is driven by the C{\sc iv} atomic doublet,  
and is visible on the blue side of most strong absorption transitions. Similar effects from the 
doublets of O{\sc vi}, Si{\sc iv}, or N{\sc v}, however, seem to not be present. Our results confirm 
that radiation pressure is a prevalent mechanism for accelerating outflows in quasars. 

\end{abstract}

\section{Introduction}

   Supermassive black holes, with masses as large as billions of times 
the mass of the Sun, inhabit the center of most, if not all, massive galaxies in the 
Universe. Despite their small size compared to that of their hosts, these 
formidable objects can determine the fate of the whole galaxy 
\citep{Furlanetto2001,Haiman2006,Scannapieco2004}, as well as 
contributing to the metal enrichment of the distant intergalactic medium \citep{Cavaliere2002,
Levine2005}. 

   When the supermassive black holes grow by undergoing episodes of gas accretion,  
they release large amounts of energy and become detectable as active galactic 
nuclei (AGNs). One of their brightest phases is represented by the so-called quasars,  
and the energy release, referred to as AGN or quasar feedback, 
results in powerful winds that can expel material out of the host galaxy  
\citep[see the reviews by][]{King2015,Harrison2017}.  

   The origin of quasar feedback may reside in the outflows launched from the innermost   
regions around the black holes, and that can reach velocities of up to a $10 - 20\,$\% of 
the speed of light \citep{Moe2009,Harrison2018}. However, probing the link between fast 
outflows and quasar feedback is challenging because the mechanisms driving these outflows 
are still unclear, and simulations cannot cover the whole range of physical scales and 
processes involved \citep{Ciotti2010,Hopkins2010,Barnes2018}. It is broadly believed that 
outflows can be 
accelerated by the pressure that radiation from the quasar exerts on the gas, but questions 
concerning the overionization, confinement, and stability of the outflowing gas, have 
been puzzling astronomers for decades \citep{Williams1972,dekool1995,Murray1995,Proga2000,Proga2007b,
Baskin2014b,Progawaters2015,Matthews2016,Waters2016,Bianchi2019}. 

    The theory of radiative acceleration in quasar outflows was initially 
developed by \cite{Mushotzky1972,Scargle1973,Burbidge1975,Braun1989,Arav1994}, and was 
inspired by that proposed for the ejection of material from massive stars by 
\cite{Milne1926} and \cite{Lucy1970}. A common feature appearing in these works is the
phenomenon known as line locking, for which absorption systems in the outflow get bound to 
a fixed relative velocity that coincides with the velocity separation of two atomic transitions, 
typically the two lines of an atomic doublet.  In detail, the outflow absorbs (or scatters) radiation from 
the quasar and thus gets accelerated. This happens for both, line and continuum quasar radiation, 
although the absorption of line radiation injects a larger boost than the continuum because the flux 
in the emission lines is the largest.  Regarding now the case of an atomic doublet, a 
region absorbing radiation from both lines of the doublet can suffer a reduction in the 
acceleration if another region between the first one and the source moves at a lower (relative) speed 
coinciding with the doublet separation. In this scenario, the region in the middle 
also absorbs radiation from the two lines of the doublet in its frame and, given the relative velocity 
between the two regions, this means that it absorbs the flux that was previously 
available for the red line of the front region. Because the back region absorbs more flux and increases its  
speed, the flux-masking disappears and the front region can again absorb the two lines and 
increase its speed accordingly, thus restoring the previous relative velocity. The acceleration of 
the front region is thus regulated by the back region, and the two regions get eventually locked in a 
metastable state where the relative velocity corresponds to the doublet separation.  

     The observation of line-locking signatures of a given species in the spectra of quasars 
can therefore confirm the presence of radiative acceleration in the outflows. 
These observations, however, are not easy, due to the complexity of the outflow absorption 
features, usually broad, and also the difficulties in distinguishing line 
locking from randomly located absorption components \citep[e.g.,][]{Korista1993}. 

     Absorption features due to outflows were first identified as strong and broad absorption lines  
(BALs) from high-ionization species such as C{\sc iv}, Si{\sc iv} or N{\sc v}, blueshifted 
from the AGN rest-frame position \citep{Lynds1967}. Nowadays, outflow features are observed in 
a variety of widths and depths, and line locking from high-ionization species is suggested in a small 
number of them: \cite{Foltz1987} and \cite{Srianand2002} reported line-locking features in the spectra 
of one quasar each, and other tentative detections in a few quasars were presented in, e.g., 
\cite{Srianand2000,Ganguly2003,Gupta2003,North2006,Lu2018}.  Furthermore, line locking 
was also detected by \cite{Bowler2014}, who analyzed composite spectra of narrow 
($\le 200\,{\rm km\,s^{-1}}$) C{\sc iv} absorption lines in the spectra of quasars.  These authors 
analyzed narrow absorbers in quasars containing BALs, but they did not use the BALs themselves,  
in order to be able to resolve the C{\sc iv} doublets and detect the line-locking signatures. 

       Although the total width of the outflow troughs can be as large as   
$\sim 30\,000\,{\rm km\,s^{-1}}$, they sometimes consist in the superposition of narrower features, 
of the order of a few hundreds of ${\rm km\,s^{-1}}$, the so-called narrow absorption lines 
\citep[NALs; e.g.,][]{Gabel2006,Lu2018b,Lu2018}.  Therefore, it should be possible to 
obtain the \textit{narrow} outflow absorption spectrum if one centers the broad troughs at a 
characteristic position when performing the composite spectra \citep{Perrotta2018}. 
Given this consideration, we conduct here a search for the line-locking signature in 36 composite 
spectra of broad ($>450\,{\rm km\,s^{-1}}$)  C{\sc iv} absorption line systems (including BALs)
representing outflows, and test the prevalence of line locking for multiple properties of the outflows, the 
quasars, and different atomic species. We assess and present the physical properties of the 
outflows inferred from the analysis of the composite spectra in our companion paper, \cite{Masribas2019}.

     In \S~\ref{sec:data} and \S~\ref{sec:methods} we detail the data and methods, respectively, and 
in \S~\ref{sec:dissection} we analyze one of our composite outflow spectra in detail.  The 
C{\sc iv}, O{\sc vi} and N{\sc v} line-locking features are assessed in \S~\ref{sec:locking}. We  
discuss our findings in \S~\ref{sec:discussion}, before concluding in \S~\ref{sec:conclusions}.

We assume a flat $\Lambda$CDM cosmology with the parameter values from \cite{Planck2015}.
  
\section{Data}\label{sec:data}

We detail in \S~\ref{sec:qsodata} the quasar spectra used in our calculations, 
and present the atomic data for the analysis of the absorption lines in \S~\ref{sec:lines}.

\begin{figure}\center 
\includegraphics[width=0.47\textwidth]{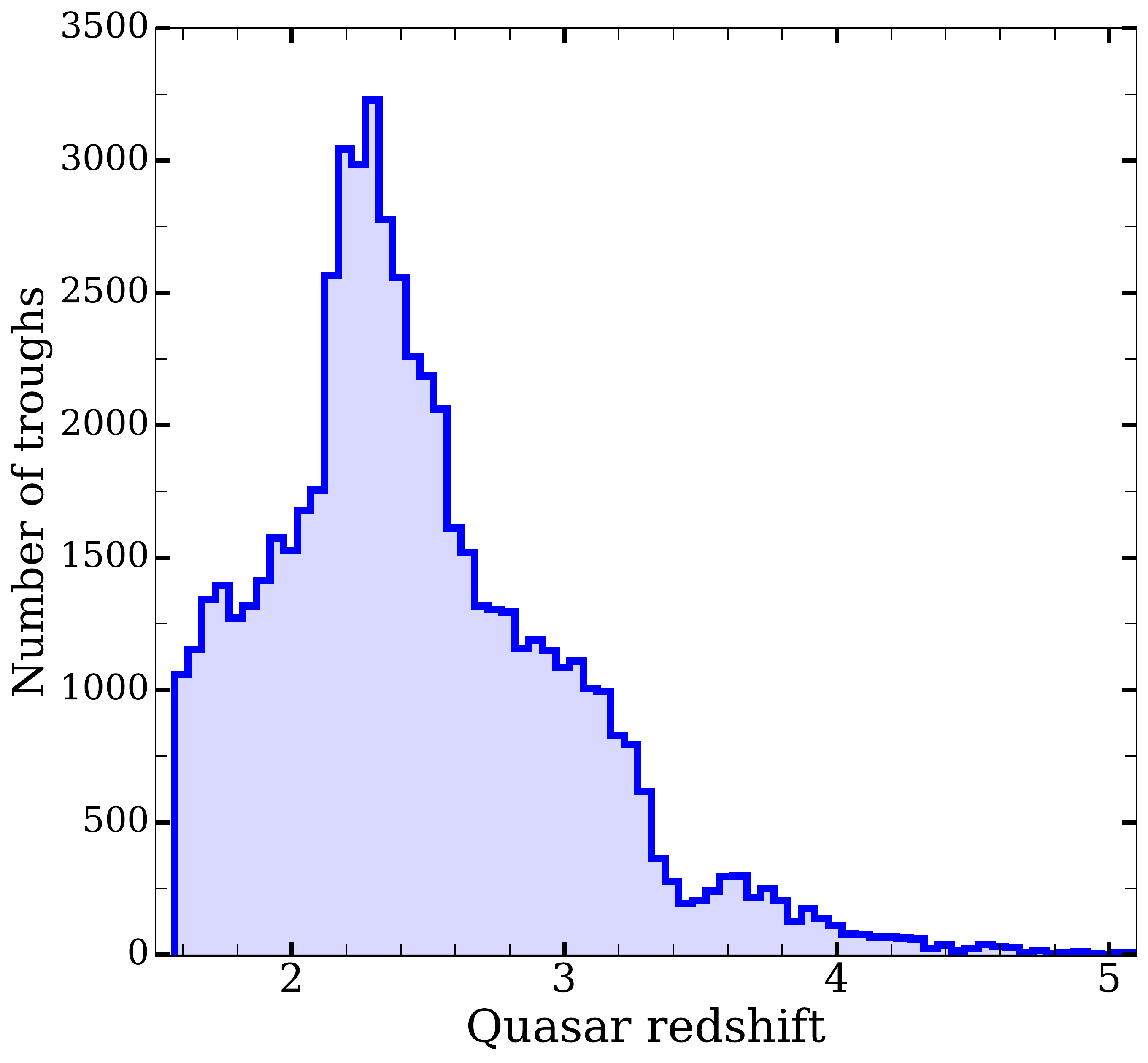}
\caption{Redshift distribution of the DR12Q quasars containing broad absorption lines, 
covering the range $1.5\lesssim z \lesssim 5.5$.} 
\label{fig:zdistr}
\end{figure}

\begin{figure}\center 
\includegraphics[width=0.48\textwidth]{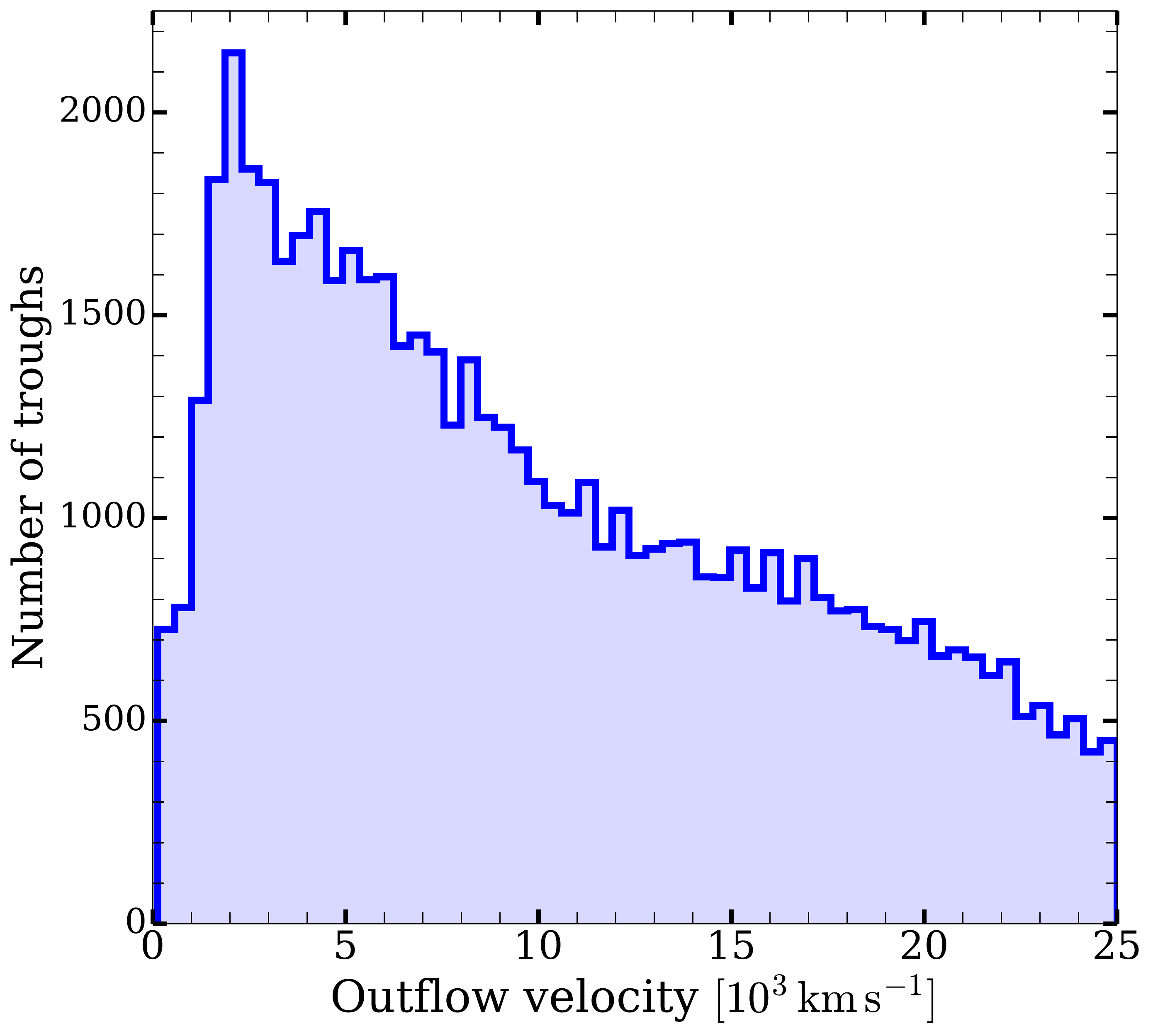}
\includegraphics[width=0.48\textwidth]{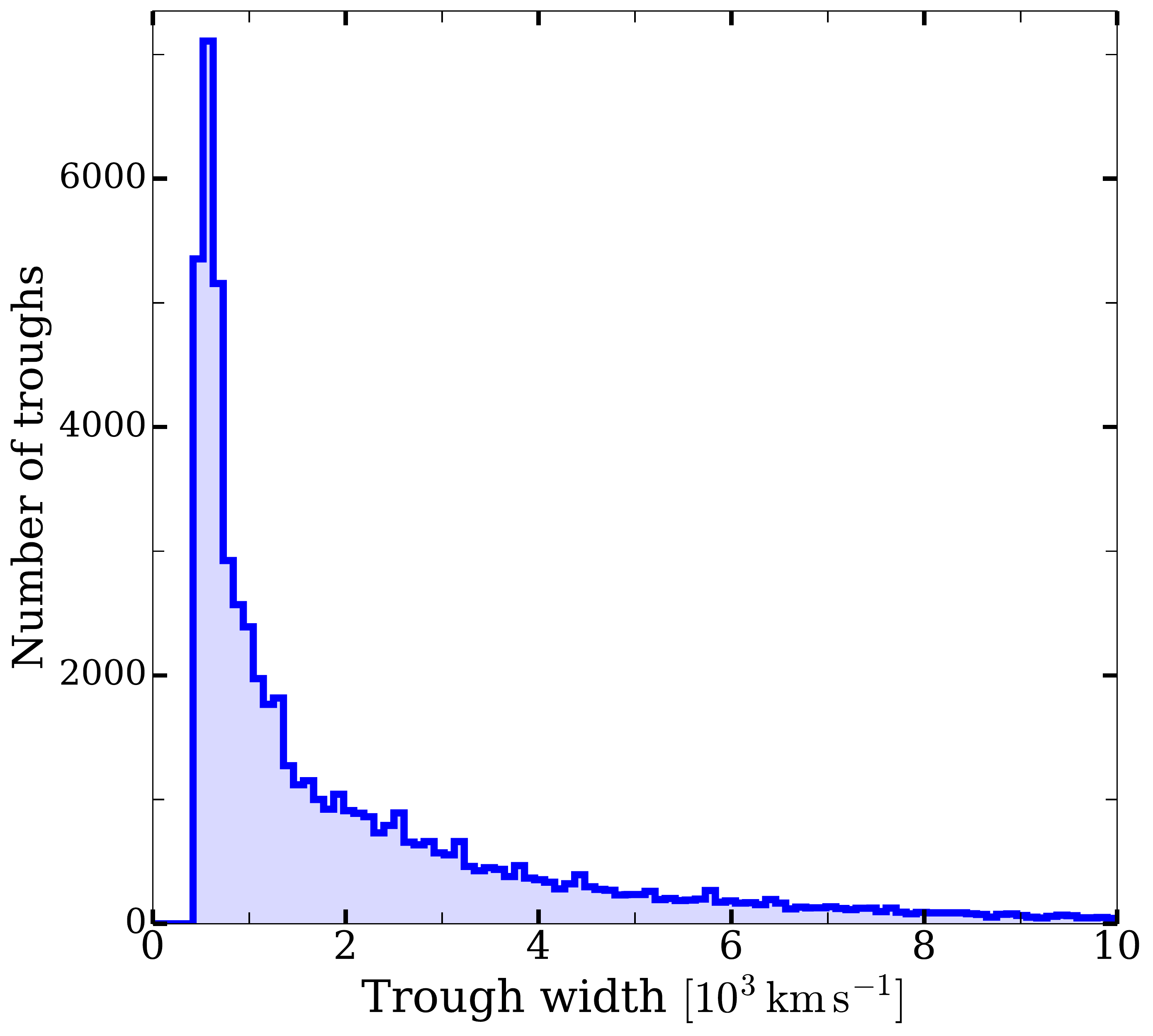}
\caption{\textit{Top panel:} Distribution of outflow velocities, with a minimum value of 
$\sim 200\;\rm{km}\,\rm{s}^{-1}$ and a peak at $\sim 2000\;\rm{km}\,\rm{s}^{-1}$, decreasing toward 
higher velocities.
\textit{Bottom panel:} Distribution of trough widths, peaking at around $700\;\rm{km}\,\rm{s}^{-1}$, and 
decreasing rapidly toward larger widths. The distribution spans from $\sim 300\;\rm{km}\,\rm{s}^{-1}$ to 
$25\,000\;\rm{km}\,\rm{s}^{-1}$, but we only plot values below $10\,000\;\rm{km}\,\rm{s}^{-1}$ due to the 
small number of objects above this threshold.}
\label{fig:veldistr}
\end{figure}

\subsection{Quasar Spectra}\label{sec:qsodata}

   We use the $297\,301$ quasar spectra in the twelfth data release of the SDSS-III/BOSS 
\citep{Eisenstein2011,Dawson2013} quasar catalog, DR12Q \citep{Paris2016}. 
The characteristics of the SDSS telescope 
and camera are detailed in \cite{Gunn1998,Gunn2006} and \cite{Ross2012}, and the 
SDSS/BOSS spectrographs in \cite{Smee2013}.  

    We also use the DR12QBAL catalog, 
which compiles the individual properties of the 
C{\sc iv} absorption troughs with widths $>450\,{\rm km\,s^{-1}}$ detected in the spectra of 
the DR12Q quasars.  The DR12QBAL catalog contains $63\,549$ troughs  with measurements   
of  their position, width and maximum depth. The position of each trough was set at the velocity 
distance from the quasar where the flux within the absorption trough is the minimum, and the trough width 
was defined as the velocity range in which the measured normalized quasar flux density 
(i.e., the flux transmission) is lower than 0.9 \citep{Paris2012,Paris2016}. We remove 
from further analysis $3\,677$ objects ($\simeq 83\%$ of them with velocity 
offset from the quasar below $v \sim 400\,{\rm km\, s^{-1}}$). In the catalog, the position of the minimum 
flux within these troughs is quoted outside the boundaries of the troughs.  The composite spectrum of 
these discarded troughs does not show significant absorption features, 
indicating that they are false positives and/or severely affected by incorrect velocity 
offset measurements or noise, and their inclusion in the samples would simply increase 
the noise at low outflow velocities. Our final sample thus consists of $59\,872$ absorption troughs. 

   Figure \ref{fig:zdistr} shows the redshift distribution of the DR12Q quasars containing troughs, which spans the range $1.5\leq z \leq 5.5$, although only a small number of 
objects have values above $z \sim 3.3$. The distribution peaks at $z\sim2.4$, similar to its mean value of $\bar z =2.46$, and decreases rapidly toward both lower and higher values.   
The {\it upper panel} in Figure \ref{fig:veldistr} shows the distribution of the velocity offset of the troughs 
from the quasars, i.e., the outflow velocities, for the final trough sample. The distribution peaks at around 
$2\,000\,{\rm km\, s^{-1}}$ and decreases toward higher velocities up to $25\,000\,{\rm km\, s^{-1}}$, with a slight change of slope that flattens beyond $\sim 12\,000\,{\rm km\, s^{-1}}$. For outflow velocities smaller than that of the peak, the decrease in the number of troughs is much steeper, 
flattening at velocities below $\sim 1\,000\,{\rm km\, s^{-1}}$. Overall, this 
distribution is similar to that obtained by \cite{Nestor2008}, who searched 
for C{\sc iv} troughs in the quasar spectra of the fourth SDSS data release. The {\it lower panel} in Figure \ref{fig:veldistr} displays the distribution of trough widths, which peaks at $\sim 700\,{\rm km\,s^{-1}}$ and decreases rapidly toward higher and lower values. The plot covers the interval $\simeq 300 - 10\,000\,{\rm km\,s^{-1}}$, although a small number of troughs have widths of up to $\sim 25\,000\,{\rm km\,s^{-1}}$.

\subsection{Absorption Lines}\label{sec:lines}

   Table \ref{ta:lines} presents vacuum atomic data in the rest frame between 
$500\,{\rm \AA}$ and $3\,000\,{\rm \AA}$, relevant for the analysis of outflow 
absorption lines. We have built this catalog by compiling data from the database {\sc linetools}\footnote{\url{https://linetools.readthedocs.io/en/latest/}} 
\citep{Linetools}, as well as that of the National Institute of Standards and Technology 
\citep[NIST\footnote{\url{https://www.nist.gov/pml/atomic-spectra-database}};][]{Nist}. 
This list includes a number of strong absorption lines from species that 
are typically not observed in interstellar or intergalactic gas studies, but that 
may arise under the extreme conditions of high temperature, density, and strong radiation 
fields present in quasar outflows. Many of these lines originate from excited meta-stable states 
above the ground level. However, this set of lines is not complete, and the interested 
reader is referred to NIST for the full spectroscopic data. The {\it first column} in Table \ref{ta:lines} 
denotes each transition, and the {\it second} one its line center in 
${\rm \AA}$ngstroms (in the vacuum). The corresponding oscillator strengths 
are shown in the {\it third column}, and the reference source in the {\it fourth} one.

     This table is publicly available in digital format at \url{https://github.com/lluism/BALs}, and 
the access is open for the community to contribute to its update, correction, and/or extension 
to make it useful for future studies.

\section{Methods}\label{sec:methods}

We detail the calculation of the mean quasar spectrum in \S~\ref{sec:cont}, and in 
\S~\ref{sec:stack} that of the outflow composite spectrum. Section \S~\ref{sec:eqw} describes the 
method for modeling the absorption lines.

\subsection{Mean Quasar Spectrum Calculation}\label{sec:cont}

We calculate a weighted-mean quasar spectrum that will 
be used to normalize the spectra for the  computation of the outflow composite spectrum, 
following our procedure in \cite{Masribas2016c}. Below, we summarize the main steps of 
this calculation, and refer the interested reader to the aforementioned work for details. 

In our calculations, we do not consider those spectral pixels affected by the skylines 
reported by \cite{Natalie2013}. Furthermore, we always correct the flux in the 
Lyman-alpha forest of the quasar spectra for the redshift-dependent average absorption 
in this region, using the analytical formula proposed by \citealt{Faucher2008} 
\citep[Eq.~7 in][]{Masribas2016c}. 
For the computation of the mean spectrum we use all the spectra in the DR12Q catalog. 
This is because the coincident position of troughs in the quasar spectra yields an 
undesired average broad absorption feature next to the quasar emission lines if only 
quasars with broad troughs are considered. 

The calculation of the mean quasar spectrum is performed as follows: we shift the 
observed spectra to the rest-frame position of the quasars, considering the visual inspection quasar redshift values in DR12Q, and rebin

\newpage
\begin{ThreePartTable}\center
\begin{TableNotes}
\footnotesize

\item {Transitions arising from excited metastable states are denoted by 
       an asterisk.}
\item {References: (1) {\sc linetools} (2) NIST. }

\end{TableNotes}

\begin{center}
\begin{longtable}{c c c c} 
\caption*{Atomic Data} \\ 

\label{ta:lines}
Transition & $\lambda\,{\rm (\AA)}$ & $f$ & Reference
\\
\hline \\
\endfirsthead

\caption*{-- Continued} \\ 
 Transition & $\lambda\,{\rm (\AA)}$  & $f$ & Reference \\ \hline \\
\endhead
\\

\insertTableNotes  
\endlastfoot

O{\sc iii} $\lambda507$ \dotfill & $507.3910$ & $0.18500$ & $1$\\
N{\sc ii} $\lambda533$ \dotfill & $533.5099$ & $0.29899$ & $1$\\
O{\sc ii} $\lambda539$ \dotfill & $539.0855$ & $0.06520$ & $1$\\
Ne{\sc iv} $\lambda541$ \dotfill & $541.1270$ & $0.03900$ & $1$\\
Ne{\sc iv} $\lambda542$ \dotfill & $542.0730$ & $0.07789$ & $1$\\
C{\sc ii} $\lambda543$ \dotfill & $543.2570$ & $0.03489$ & $1$\\
Ne{\sc iv} $\lambda543$ \dotfill & $543.8910$ & $0.11599$ & $1$\\
O{\sc iv} $\lambda553$ \dotfill & $553.3300$ & $0.11200$ & $1$\\
O{\sc iv} $\lambda554$ \dotfill & $554.0750$ & $0.22400$ & $1$\\
Ne{\sc vi} $\lambda558$ \dotfill & $558.5900$ & $0.09070$ & $1$\\
Ne{\sc vi} $\lambda558$ \dotfill & $558.5900$ & $0.09070$ & $1$\\
C{\sc ii} $\lambda560$ \dotfill & $560.2394$ & $0.05710$ & $1$\\
Ne{\sc v} $\lambda568$ \dotfill & $568.4200$ & $0.09279$ & $1$\\
He{\sc i} $\lambda584$ \dotfill & $584.3300$ & $0.27625$ & $2$\\
C{\sc ii} $\lambda594$ \dotfill & $594.8000$ & $0.11699$ & $1$\\
O{\sc iv} $\lambda608$ \dotfill & $608.3980$ & $0.06700$ & $1$\\
Mg{\sc x} $\lambda609$ \dotfill & $609.7900$ & $0.08420$ & $1$\\
Mg{\sc x} $\lambda624$ \dotfill & $624.9500$ & $0.04100$ & $1$\\
O{\sc iv} $\lambda625$ \dotfill & $625.0000$ & $0.12500$ & $2$\\
O{\sc v} $\lambda629$ \dotfill & $629.7300$ & $0.51499$ & $1$\\
S{\sc ii} $\lambda641$ \dotfill & $641.7670$ & $0.25999$ & $1$\\
S{\sc iv} $\lambda657$ \dotfill & $657.3280$ & $1.13000$ & $2$\\
S{\sc iv} $\lambda661$ \dotfill & $661.4430$ & $1.02000$ & $2$\\
S{\sc iii} $\lambda680$ \dotfill & $680.6800$ & $1.38000$ & $2$\\
S{\sc iii} $\lambda681$ \dotfill & $681.4700$ & $0.06830$ & $1$\\
S{\sc iii} $\lambda683$ \dotfill & $683.5860$ & $1.34000$ & $2$\\
N{\sc iii} $\lambda684$ \dotfill & $684.9960$ & $0.13500$ & $1$\\
N{\sc iii} $\lambda685$ \dotfill & $685.8200$ & $0.32000$ & $2$\\
S{\sc iii} $\lambda698$ \dotfill & $698.7310$ & $0.78299$ & $1$\\
O{\sc iii} $\lambda702$ \dotfill & $702.3320$ & $0.13699$ & $1$\\
S{\sc iii} $\lambda724$ \dotfill & $724.2890$ & $0.35199$ & $1$\\
S{\sc iv} $\lambda744$ \dotfill & $744.9070$ & $0.25099$ & $1$\\
S{\sc iv} $\lambda748$ \dotfill & $748.4000$ & $0.50000$ & $1$\\
N{\sc iii} $\lambda763$ \dotfill & $763.3400$ & $0.08200$ & $1$\\
S{\sc ii} $\lambda763$ \dotfill & $763.6570$ & $0.39800$ & $1$\\
S{\sc ii} $\lambda764$ \dotfill & $764.4200$ & $0.79500$ & $1$\\
N{\sc iv} $\lambda765$ \dotfill & $765.1480$ & $0.61599$ & $1$\\
S{\sc ii} $\lambda765$ \dotfill & $765.6930$ & $1.19000$ & $1$\\
Ne{\sc viii} $\lambda770$ \dotfill & $770.4090$ & $0.10300$ & $1$\\
Ne{\sc viii} $\lambda780$ \dotfill & $780.3240$ & $0.05050$ & $1$\\
S{\sc v} $\lambda786$ \dotfill & $786.4700$ & $1.36000$ & $2$\\
O{\sc iv} $\lambda787$ \dotfill & $787.7110$ & $0.11100$ & $1$\\
O{\sc i} $\lambda791$ \dotfill & $791.9732$ & $0.04639$ & $1$\\
S{\sc iv} $\lambda809$ \dotfill & $809.6680$ & $0.10400$ & $1$\\
S{\sc iv} $\lambda815$ \dotfill & $815.9450$ & $0.08500$ & $2$\\
O{\sc ii} $\lambda832$ \dotfill & $832.7572$ & $0.04439$ & $1$\\
O{\sc iii} $\lambda832$ \dotfill & $832.9270$ & $0.10700$ & $1$\\
O{\sc ii} $\lambda833$ \dotfill & $833.3294$ & $0.08860$ & $1$\\
O{\sc ii} $\lambda834$ \dotfill & $834.4655$ & $0.13199$ & $1$\\
Fe{\sc iii} $\lambda844$ \dotfill & $844.2880$ & $0.06840$ & $1$\\
C{\sc ii} $\lambda858$ \dotfill & $858.0918$ & $0.01300$ & $1$\\
Fe{\sc iii} $\lambda859$ \dotfill & $859.7230$ & $0.11500$ & $1$\\
O{\sc i} $\lambda877a$ \dotfill & $877.7983$ & $0.01970$ & $1$\\
O{\sc i} $\lambda877b$ \dotfill & $877.8787$ & $0.05889$ & $1$\\
Si{\sc ii} $\lambda889$ \dotfill & $889.7228$ & $0.04340$ & $1$\\
S{\sc ii} $\lambda906$ \dotfill & $906.8850$ & $0.21000$ & $1$\\
P{\sc iii} $\lambda913$ \dotfill & $913.9683$ & $0.20300$ & $1$\\
N{\sc ii} $\lambda915$ \dotfill & $915.6131$ & $0.15900$ & $1$\\
P{\sc iii} $\lambda917$ \dotfill & $917.1178$ & $0.40400$ & $1$\\
S{\sc vi} $\lambda933$ \dotfill & $933.3780$ & $0.43700$ & $1$\\
H{\sc i} $\lambda937$ \dotfill & $937.8034$ & $0.00780$ & $1$\\
S{\sc vi} $\lambda944$ \dotfill & $944.5230$ & $0.21500$ & $1$\\
C{\sc i} $\lambda945$ \dotfill & $945.1910$ & $0.15200$ & $1$\\
O{\sc i} $\lambda948$ \dotfill & $948.6855$ & $0.00631$ & $1$\\
H{\sc i} $\lambda949$ \dotfill & $949.7430$ & $0.01395$ & $1$\\
P{\sc iv} $\lambda950$ \dotfill & $950.6600$ & $1.60000$ & $2$\\
N{\sc i} $\lambda963$ \dotfill & $963.9903$ & $0.01240$ & $1$\\
N{\sc i} $\lambda964$ \dotfill & $964.6256$ & $0.00790$ & $1$\\
H{\sc i} $\lambda972$ \dotfill & $972.5367$ & $0.02901$ & $1$\\
C{\sc iii} $\lambda977$ \dotfill & $977.0201$ & $0.75700$ & $1$\\
O{\sc i} $\lambda988$ \dotfill & $988.7734$ & $0.04650$ & $1$\\
N{\sc iii} $\lambda989$ \dotfill & $989.7990$ & $0.12300$ & $1$\\
Si{\sc ii} $\lambda989$ \dotfill & $989.8731$ & $0.17100$ & $1$\\
S{\sc iii} $\lambda1012$ \dotfill & $1012.495$ & $0.04380$ & $1$\\
Si{\sc ii} $\lambda1020$ \dotfill & $1020.698$ & $0.01680$ & $1$\\
H{\sc i} $\lambda1025$ \dotfill & $1025.722$ & $0.07914$ & $1$\\
O{\sc i} $\lambda1027$ \dotfill & $1027.431$ & $0.01962$ & $2$\\
O{\sc i} $\lambda1028$ \dotfill & $1028.157$ & $0.02010$ & $2$\\
O{\sc vi} $\lambda1031$ \dotfill & $1031.926$ & $0.13250$ & $1$\\
C{\sc ii} $\lambda1036$ \dotfill & $1036.336$ & $0.11800$ & $1$\\
C{\sc ii} $\lambda1037$ \dotfill & $1037.018$ & $0.11800$ & $2$\\ 
O{\sc vi} $\lambda1037$ \dotfill & $1037.616$ & $0.06580$ & $1$\\
O{\sc i} $\lambda1039$ \dotfill & $1039.230$ & $0.00907$ & $1$\\
Fe{\sc ii} $\lambda1055$ \dotfill & $1055.261$ & $0.00750$ & $1$\\
S{\sc iv} $\lambda1062$ \dotfill & $1062.664$ & $0.04940$ & $1$\\
S{\sc iv}* $\lambda1073$ \dotfill & $1072.962$ & $0.04200$ & $2$\\
Fe{\sc ii} $\lambda1081$ \dotfill & $1081.874$ & $0.01260$ & $1$\\
N{\sc ii} $\lambda1083$ \dotfill & $1083.993$ & $0.11100$ & $1$\\
Fe{\sc ii} $\lambda1096$ \dotfill & $1096.876$ & $0.03200$ & $1$\\
Fe{\sc ii} $\lambda1112$ \dotfill & $1112.048$ & $0.00620$ & $1$\\
C{\sc i} $\lambda1112$ \dotfill & $1112.269$ & $0.01610$ & $1$\\
P{\sc v} $\lambda1117$ \dotfill & $1117.977$ & $0.47200$ & $1$\\
Fe{\sc ii} $\lambda1121$ \dotfill & $1121.974$ & $0.02020$ & $1$\\
C{\sc i} $\lambda1122$ \dotfill & $1122.437$ & $0.00511$ & $1$\\
Fe{\sc iii} $\lambda1122$ \dotfill & $1122.524$ & $0.05440$ & $1$\\
Fe{\sc ii} $\lambda1125$ \dotfill & $1125.447$ & $0.01600$ & $1$\\
P{\sc v} $\lambda1128$ \dotfill & $1128.007$ & $0.23300$ & $1$\\
C{\sc i} $\lambda1129$ \dotfill & $1129.195$ & $0.00771$ & $1$\\
Fe{\sc ii} $\lambda1133$ \dotfill & $1133.665$ & $0.00550$ & $1$\\
N{\sc i} $\lambda1134a$ \dotfill & $1134.165$ & $0.01460$ & $1$\\
N{\sc i} $\lambda1134b$ \dotfill & $1134.414$ & $0.02870$ & $1$\\
N{\sc i} $\lambda1134c$ \dotfill & $1134.980$ & $0.04160$ & $1$\\
Fe{\sc ii} $\lambda1143$ \dotfill & $1143.226$ & $0.01920$ & $1$\\
Fe{\sc ii} $\lambda1144$ \dotfill & $1144.937$ & $0.08300$ & $1$\\
C{\sc i} $\lambda1158$ \dotfill & $1158.324$ & $0.00655$ & $1$\\
C{\sc iii}* $\lambda1175$ \dotfill & $1175.260$ & $0.27240$ & $2$\\
S{\sc iii} $\lambda1190$ \dotfill & $1190.203$ & $0.02370$ & $1$\\
Si{\sc ii} $\lambda1190$ \dotfill & $1190.415$ & $0.29200$ & $1$\\
C{\sc i} $\lambda1193$ \dotfill & $1193.030$ & $0.04090$ & $1$\\
Si{\sc ii} $\lambda1193$ \dotfill & $1193.289$ & $0.58200$ & $1$\\
Si{\sc ii} $\lambda1194$ \dotfill & $1194.500$ & $0.73700$ & $2$\\
Mn{\sc ii} $\lambda1197$ .\dotfill & $1197.184$ & $0.21700$ & $1$\\
Mn{\sc ii} $\lambda1199$ . . .\dotfill & $1199.391$ & $0.16900$ & $1$\\
N{\sc i} $\lambda1199$ \dotfill & $1199.549$ & $0.13200$ & $1$\\
N{\sc i} $\lambda1200$ \dotfill & $1200.223$ & $0.08690$ & $1$\\
N{\sc i} $\lambda1200b$ \dotfill & $1200.709$ & $0.04320$ & $1$\\
Si{\sc iii} $\lambda1206$ \dotfill & $1206.500$ & $1.63000$ & $1$\\
H{\sc i} $\lambda1215$ \dotfill & $1215.670$ & $0.41640$ & $1$\\
N{\sc v} $\lambda1238$ \dotfill & $1238.821$ & $0.15600$ & $1$\\
N{\sc v} $\lambda1242$ \dotfill & $1242.804$ & $0.07770$ & $1$\\
S{\sc ii} $\lambda1250$ \dotfill & $1250.578$ & $0.00543$ & $1$\\
S{\sc ii} $\lambda1253$ \dotfill & $1253.805$ & $0.01090$ & $1$\\
Si{\sc ii} $\lambda1260$ \dotfill & $1260.422$ & $1.18000$ & $1$\\
Si{\sc ii} $\lambda1264$ \dotfill & $1264.730$ & $1.09000$ & $2$\\
C{\sc i} $\lambda1276$ \dotfill & $1276.482$ & $0.00589$ & $1$\\
C{\sc i} $\lambda1277$ \dotfill & $1277.245$ & $0.08530$ & $1$\\
O{\sc i} $\lambda1302$ \dotfill & $1302.168$ & $0.04800$ & $1$\\
Si{\sc ii} $\lambda1304$ \dotfill & $1304.370$ & $0.08630$ & $1$\\
Ni{\sc ii} $\lambda1317$ \dotfill & $1317.217$ & $0.14599$ & $1$\\
C{\sc i} $\lambda1328$ \dotfill & $1328.833$ & $0.07580$ & $1$\\
C{\sc ii} $\lambda1334$ \dotfill & $1334.532$ & $0.12800$ & $1$\\
C{\sc ii} $\lambda1335$ \dotfill & $1335.708$ & $0.11500$ & $2$\\
Ni{\sc ii} $\lambda1370$ \dotfill & $1370.132$ & $0.07690$ & $1$\\
Si{\sc iv} $\lambda1393$ \dotfill & $1393.755$ & $0.52399$ & $1$\\
Si{\sc iv} $\lambda1402$ \dotfill & $1402.770$ & $0.25999$ & $1$\\
Ni{\sc ii} $\lambda1454$ \dotfill & $1454.842$ & $0.03230$ & $1$\\
Si{\sc ii} $\lambda1526$ \dotfill & $1526.707$ & $0.12700$ & $1$\\
C{\sc iv} $\lambda1548$ \dotfill & $1548.204$ & $0.18999$ & $1$\\
C{\sc iv} $\lambda1550$ \dotfill & $1550.770$ & $0.09520$ & $1$\\
C{\sc i} $\lambda1560$ \dotfill & $1560.309$ & $0.07740$ & $1$\\
Si{\sc i} $\lambda1589$ \dotfill & $1589.174$ & $0.05040$ & $1$\\
Fe{\sc ii} $\lambda1608$ \dotfill & $1608.451$ & $0.05770$ & $1$\\
C{\sc i} $\lambda1656$ \dotfill & $1656.928$ & $0.14900$ & $1$\\
Al{\sc ii} $\lambda1670$ \dotfill & $1670.788$ & $1.74000$ & $1$\\
Ni{\sc ii} $\lambda1703$ \dotfill & $1703.411$ & $0.00600$ & $1$\\
Ni{\sc ii} $\lambda1709$ \dotfill & $1709.604$ & $0.03240$ & $1$\\
Ni{\sc ii} $\lambda1741$ \dotfill & $1741.553$ & $0.04270$ & $1$\\
Ni{\sc ii} $\lambda1751$ \dotfill & $1751.915$ & $0.02770$ & $1$\\
Ni{\sc ii} $\lambda1773$ \dotfill & $1773.949$ & $0.00621$ & $1$\\
Ni{\sc ii} $\lambda1804$ \dotfill & $1804.473$ & $0.00716$ & $1$\\
Mg{\sc i} $\lambda1827$ \dotfill & $1827.935$ & $0.02420$ & $1$\\
Al{\sc iii} $\lambda1854$ \dotfill & $1854.716$ & $0.57499$ & $1$\\
Al{\sc iii} $\lambda1862$ \dotfill & $1862.789$ & $0.28600$ & $1$\\
Zn{\sc ii} $\lambda2026$ \dotfill & $2026.137$ & $0.50100$ & $1$\\
Cr{\sc ii} $\lambda2056$ \dotfill & $2056.256$ & $0.10300$ & $1$\\
Cr{\sc ii} $\lambda2062$ \dotfill & $2062.234$ & $0.10499$ & $1$\\
Zn{\sc ii} $\lambda2062$ \dotfill & $2062.664$ & $0.25299$ & $1$\\
Cr{\sc ii} $\lambda2066$ \dotfill & $2066.161$ & $0.06979$ & $1$\\
Fe{\sc ii} $\lambda2344$ \dotfill & $2344.213$ & $0.11400$ & $1$\\
Fe{\sc ii} $\lambda2374$ \dotfill & $2374.461$ & $0.03130$ & $1$\\
Fe{\sc ii} $\lambda2382$ \dotfill & $2382.765$ & $0.32000$ & $1$\\
Mn{\sc ii} $\lambda2576$ \dotfill & $2576.877$ & $0.36100$ & $1$\\
Fe{\sc ii} $\lambda2586$ \dotfill & $2586.650$ & $0.06910$ & $1$\\
Mn{\sc ii} $\lambda2594$ \dotfill & $2594.499$ & $0.28000$ & $1$\\
Fe{\sc ii} $\lambda2600$ \dotfill & $2600.172$ & $0.23900$ & $1$\\
Mn{\sc ii} $\lambda2606$ \dotfill & $2606.462$ & $0.19800$ & $1$\\
Mg{\sc ii} $\lambda2796$ \dotfill & $2796.354$ & $0.61550$ & $1$\\
Mg{\sc ii} $\lambda2803$ \dotfill & $2803.531$ & $0.30580$ & $1$\\
Mg{\sc i} $\lambda2852$ \dotfill & $2852.964$ & $1.73000$ & $1$ \\

\hline  
\end{longtable}
\end{center}
\end{ThreePartTable}

the flux density and error into 
an evenly-spaced wavelength array with pixel width of $1\, {\rm \AA}$. The spectra 
are then normalized by dividing them by their corresponding mean flux in the 
wavelength windows free of emission lines $1\,300<\lambda_{\rm r}/{\rm \AA}<1\,383$ and 
$1\,408<\lambda_{\rm r}/{\rm \AA}<1\,500$. 
If more than $20\%$ of the pixels in these intervals for one spectrum are not accounted 
for due to skylines, this spectrum is not used. This constraint rejects less than $10\%$ 
of the spectra in the catalog. 

    A mean signal-to-noise ratio (S/N) is then calculated 
for each remaining spectrum $j$ as 
\begin{equation}
{\rm S/N}_j = \frac{\sum_i{f_{ij}/N_j}}{\left(\sum_i{e_{ij}}^2/N_j\right)^{1/2}} ~,
\end{equation}
where $e_{ij}$ is the uncertainty for the flux $f_{ij}$ at the pixel $i$, and $N_j$ is 
the number of pixels within the above wavelength windows intervening in the summation. 
Those spectra with S/N$<1$ are discarded from further calculations. This threshold 
eliminates around $22\%$ of the remaining spectra from the calculations. Finally, the 
mean S/N values are used for assigning to each quasar spectrum a weight of the form
\begin{equation}
w_j = {{1}\over{{{\rm S/N}_j}^{-2} + \sigma^2}} ~,
\end{equation}
where $\sigma=0.1$ limits the potential excessive contribution from spectra with very 
high S/N. In Mas-Ribas et al. 2017, we tested that small variations around this value do 
not alter our results, and we therefore use the same quantity. The final mean quasar 
spectrum is then computed as a weighted mean, using the aforementioned weight values. 

\begin{figure}\center 
\includegraphics[width=0.48\textwidth]{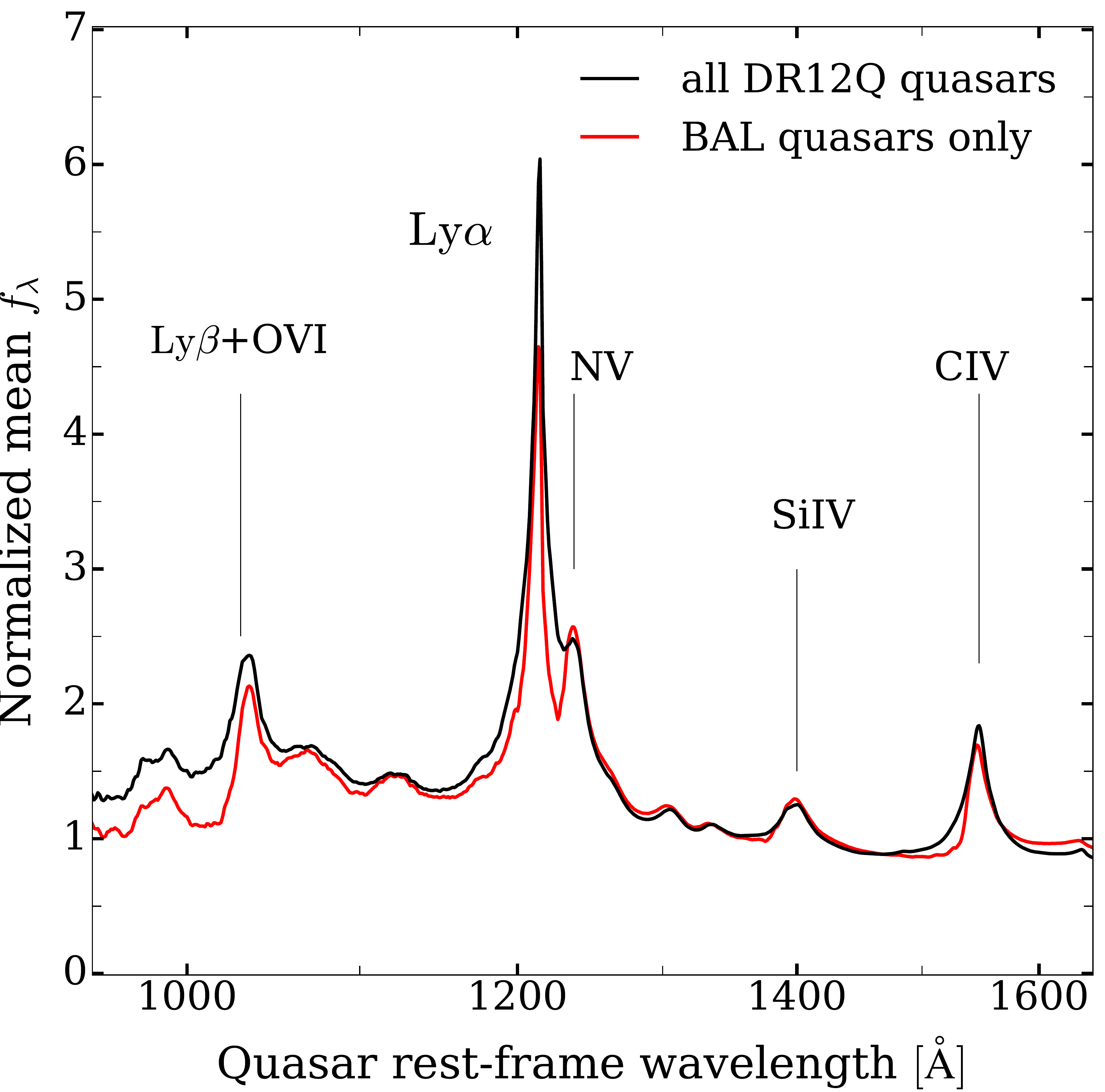}
\caption{Mean quasar spectrum considering all DR12Q quasars (\textit{black line}), and only those 
containing broad absorption troughs 
(\textit{red line}). The average outflow absorption features of the elements with highest opacities, such as 
C{\sc iv},  N{\sc v}, Ly$\alpha$ and  O{\sc vi}, are visible on the left part of the respective emission 
lines in the red spectrum. The horizontal axis is logarithmically spaced for visualization.}
\label{fig:cont}
\end{figure}

    Figure \ref{fig:cont} shows the difference between the spectrum considering all the 
DR12Q quasar spectra ({\it black line}), and only those containing troughs ({\it red 
line}). A broad absorption trough is visible to the left of the emission lines of  
species with the highest opacities, i.e., C{\sc iv},  N{\sc v}, Ly$\alpha$ and  
O{\sc vi}, in the trough-only spectrum. The use of this biased mean spectrum would result 
in the underestimation of the true line equivalent widths.
We have checked that including the quasars containing troughs to the total quasar sample for 
this calculation does not affect significantly the resulting mean quasar spectrum.   
The relative difference between the mean quasar built from the whole quasar sample  
and that from only quasars not containing outflow troughs, is $\lesssim 6\%$ at the peaks of the 
quasar emission lines. In the flatter quasar regions where the absorption troughs are found, the 
differences are $\lesssim 2\%$. A similar mean 
spectrum, except for the lack of corrections on the average absorption of the 
Lyman-alpha forest, is illustrated in Figure 3 of \cite{Masribas2016c}, and a movie 
showing the build up of the mean quasar spectrum with an increasing number of spectra,  
is publicly available at \url{https://github.com/lluism/DLA_movies}.

     We have tested the impact on our results of using an individual mean spectrum for 
every quasar spectrum, derived from the spectra of its nearest neighbors, instead of our 
mean quasar calculation. In the nearest neighbor approach, the mean spectrum for each 
quasar containing troughs is obtained by computing the mean spectrum of its forty trough-free 
nearest neighbors among the entire DR12Q catalog.  The proximity between the spectra 
is defined as the Euclidean distance in normalized flux units as  
\begin{equation}
d_{jk} = \sqrt{\sum_i (f^N_{i,j} - f^N_{i,k})^2 }~,
\end{equation} 
where $i$ are the pixels in the quasars $j$ and $k$, and $f^N_i$ is the flux once normalized by 
the average flux in the wavelength windows $1\,300<\lambda_{\rm r}/{\rm \AA}<1\,383$ and 
$1\,408<\lambda_{\rm r}/{\rm \AA}<1\,500$. We have considered only the 
spectral regions above $1\,280$ {\rm \AA}  in the quasar rest frame to avoid the impact of the 
strongest quasar emission lines \citep{Davies2018}. Using the mean spectra  
from the nearest-neighbor approach, yields outflow composite spectra slightly less noisy 
redward of the C{\sc iv} absorption doublet, compared to using the mean of all the quasars.  
This is because the individual mean, i.e., from the nearest neighbors, matches  
the quasar spectra better than the overall mean. However, the noise for the 
nearest-neighbor 
case is significantly larger at short wavelengths, especially in the Lyman-alpha forest, 
since not all the spectra in the DR12Q catalog cover the region below 
$\sim 1\,500\,{\rm \AA}$ in the quasar rest-frame. If one considers only the spectra 
that cover the entire forest region ($\gtrsim 912\,{\rm \AA}$), then the number of 
spectra and candidate neighbors is reduced, and the resulting outflow composite does not 
show a significant improvement compared to using our mean spectra. We, therefore,  
conclude that our mean quasar spectrum method is adequate for our calculations.

\subsection{Outflow Composite Spectra Calculation}\label{sec:stack}

   We compute the composite rest-frame outflow spectrum as a weighted mean, now 
only using the quasar spectra that contain troughs, and centering these spectra 
at the position of the minimum flux within the troughs as given in DR12QBAL. 
\cite{Perrotta2018} recently stacked 
outflow absorption features in high-resolution spectra, and they found that centering the 
spectra at the position of the deepest (minimum flux) C{\sc iv} component, yields the 
best alignment between C{\sc iv} and the other absorption features, especially those 
of low-ionization species. In our case, given the low resolution of the BOSS spectra, 
the position of the minimum flux can be determined with higher accuracy than the 
limits of the troughs, or other points, which in turn results in more precise 
(narrower) absorption profiles. Indeed, our stacks 
do not resolve the two lines of the C{\sc iv} doublet, and show much 
broader absorption features, if we center the spectra at the centroid of the 
troughs, or at the minimum or maximum trough limits. This test suggests that the 
minimum flux position mostly has a physical origin, and that the effect of 
noise is small. We elaborate further on the 
impact of choosing the minimum-flux position on our results in \S~\ref{sec:really}.

   The redshift of the troughs is computed using the velocity offset from the quasar (i.e., 
from the C{\sc iv} quasar emission lines) 
given in the DR12QBAL catalog, via the expression  
\begin{equation}\label{eq:zvl}
\frac{v}{c} = \frac{(1+z_{\rm q})^2 - (1+z_{\rm t})^2}{(1+z_{\rm q})^2 + (1+z_{\rm t})^2}~.
\end{equation}
Here, $v$ and $c$ are the trough position and the speed of light,  
respectively, and $z_{\rm t}$ and $z_{\rm q}$ denote the redshifts of the trough 
and the quasar, respectively. We then rebin each outflow spectrum into an evenly-spaced 
array of pixel width ${\rm d}\lambda = 0.3\,{\rm \AA}$, similar to the maximum BOSS 
pixel resolution ($\simeq 69\,{\rm km\,s^{-1}}$ in velocity space), and normalize them 
with the corresponding normalization 
factor obtained in the mean spectrum calculation.
Next, the normalized spectra are divided by the mean quasar spectrum, shifted and rebined 
to the new outflow wavelength array, to eliminate the quasar footprint and obtain the individual 
outflow absorption spectra. The final outflow composite spectrum results from computing the 
weighted mean of these individual outflow absorption spectra, where the weighting factors are 
the weights obtained for every spectrum in the mean quasar calculation. At the end of these 
calculations, the composite spectra present a residual offset of $\approx 200\,{\rm km\,^{-1}}$ 
from the theoretical position expected for the absorption lines. This is because 
the position of the absorption trough is measured from the peak of the C{\sc iv} quasar 
emission line, that is generally found at a rest-frame wavelength of 
$\approx 1\,549.4\,{\rm \AA}$, while the minimum flux within the trough is denoted by 
the blue line of the C{\sc iv} doublet at $\approx 1\,548.2\,{\rm \AA}$, thus separated from the 
emission peak by $\approx 250\,{\rm km\,^{-1}}$. This separation is consistent with the 
observed offset considering our flux pixel resolution of $\sim 69\,{\rm km\,s^{-1}}$. We 
simply correct for this offset by shifting our composite spectra to the theoretical position 
of the absorption lines.

\begin{table}\center    
	\begin{center}
	\caption{Outflow samples}	\label{ta:subsample}
	\begin{threeparttable}
		\begin{tabular}{lrcr} 
		\hline 
		 Selection criteria		     &Mean		&~ $n\,{\rm (\AA)}$\,\tnote{($1$)} $\;$   &No. troughs	 \\ 
         all 									&			&$0.437\pm0.029$	&$59\,872$ \\ \hline
         
		\multicolumn{4}{c}{Outflow velocity (${\rm km\,s^{-1}}$)}	     \\ 
        $v  < 200$   	             &$139$	     &$0.335\pm0.064$	 &$185$      \\
        $v  < 350$                   &$264$	    		 &$0.331\pm0.045$	 &$3\,167$     \\
        $v  < 650$   	             &$315$	     &$0.254\pm0.040$	 &$4\,355$     \\
        $350 \leq v  < 650$          &$452$	     &$0.278\pm0.038$	 &$1\,188$     \\
        $650 \leq v  < 1\,500$         &$1\,122$   &$0.291\pm0.037$	 &$2\,278$     \\
        $1\,500 \leq v  < 3\,000$        &$2\,233$   &$0.387\pm0.081$	 &$6\,706$     \\
        $3\,000 \leq v  < 5\,000$        &$3\,980$   &$0.444\pm0.095$	 &$7\,727$     \\
        $5\,000 \leq v  < 8\,000$        &$6\,426$   &$0.462\pm0.031$	 &$10\,183$  \\
        $8\,000 \leq v  < 13\,000$     &$10\,330$  &$0.439\pm0.050$	 &$12\,488$  \\
        $13\,000 \leq v  < 17\,500$  &$15\,208$	 &$0.354\pm0.078$	 &$9\,075$     \\
        $v \geq 17\,500$   			 &$20\,867$	 &$0.302\pm0.138$	 &$10\,737$  \\ \hline
        
		\multicolumn{4}{c}{Trough width (${\rm km\,s^{-1}}$)}			   \\ 
        $v  < 560$   				&$508$	   &$0.227\pm0.045$	 &$11\,250$    \\
        $560 \leq v  < 708$  		&$629$	   &$0.264\pm0.094$	 &$8\,420$       \\
        $708 \leq v  < 1\,260$   	&$951$	   &$0.305\pm0.149$	 &$13\,488$    \\
        $1\,260 \leq v  < 2\,240$   &$1\,704$  &$0.381\pm0.057$	 &$10\,266$    \\
        $v  \geq 2\,240$   		&$5\,179$  &$0.560\pm0.056$	 &$20\,125$    \\ \hline

		\multicolumn{4}{c}{Trough min. velocity (${\rm km\,s^{-1}}$)}		\\ 
        $v  < 300$   			 &$128$		&$0.423\pm0.048$	&$2\,574$   \\
        $300 \leq v  < 2\,300$  	 &$1\,236$	&$0.431\pm0.075$	&$10\,713$  \\
        $2\,300 \leq v  < 5\,300$     &$3\,694$	&$0.381\pm0.132$	&$10\,661$  \\
        $5\,300 \leq v < 9\,000$   	 &$7\,113$	&$0.352\pm0.032$	&$10\,356$  \\
        $9\,000 \leq v < 13\,500$    &$11\,183$	&$0.400\pm0.082$	&$10\,570$  \\
        $13\,500 \leq v < 18\,500$  &$15\,953$	&$0.409\pm0.054$	&$10\,059$  \\
        $v  \geq 18\,500$   		 &$21\,186$	&$0.463\pm0.089$	&$8\,630$   \\ \hline

		\multicolumn{4}{c}{Quasar redshift}		\\ 
        $z  < 1.95$   			&$1.77$	&$0.449\pm0.093$	&$10\,580$             \\
        $1.95 \leq z  < 2.20$  	&$2.09$	&$0.455\pm0.047$	&$10\,584$             \\
        $2.20 \leq z  < 2.37$   	&$2.28$	&$0.455\pm0.042$	&$10\,812$             \\
        $2.37 \leq z < 2.60$   	&$2.48$	&$0.414\pm0.066$	&$10\,591$             \\
        $2.60 \leq z < 3.00$   	&$2.79$	&$0.452\pm0.009$	&$10\,880$             \\
        $z  \geq 3.00$   		&$3.38$	&$0.461\pm0.159$	&$10\,098$             \\ \hline

		\multicolumn{4}{c}{Quasar magnitude (mag)}		\\ 
        $-25.0 \leq M_i  < -22.0$  	&$-24.6$	&$0.522\pm0.191$	&$10\,972$    \\
        $-25.5 \leq M_i  < -25.0$  	&$-25.3$	&$0.482\pm0.100$	&$10\,912$    \\
        $-26.0 \leq M_i  < -25.5$   &$-25.8$	&$0.428\pm0.061$	&$12\,932$    \\
        $-26.5 \leq M_i < -26.0$   	&$-26.2$	&$0.415\pm0.040$	&$12\,020$    \\
        $-27.0 \leq M_i < -26.5$   	&$-26.7$	&$0.387\pm0.037$	&$8\,949$     \\
        $-30.0  \leq M_i < -27.0$   &$-27.5$	&$0.383\pm0.074$	&$7\,749$     \\ \hline
		
		\end{tabular}	
		\begin{tablenotes}
			\item[($1$)] This parameter denotes the values of the intercept in Eq.~\ref{eq:aeq} 
			and is an indicator of the width of the absorption lines in the composite spectra. 
		\end{tablenotes}
	\end{threeparttable}
	\end{center}
\end{table}

  We build 36 outflow composite spectra, drawn from subsamples of the overall 
trough catalog, all computed as described above. We divide the outflow velocity 
range in nine bins, and consider two additional subsamples  
for low velocities. The trough width range is divided in five samples, and the 
trough minimum velocity in seven. The quasar redshift and magnitude ranges are 
divided in six bins each. The choice for the subsamples and their number 
is performed with the aim to divide the data in enough bins to study the dependences 
on outflow and quasar parameters in our companion work, while containing a similar number 
of troughs in each bin to reach similar signal-to-noise levels. When effects are observed at 
specific values (i.e., at low outflow velocities), we create additional bins to study them in more 
detail, even though this reduces the number of troughs. Together with the stack of all troughs 
in the catalog, these subsamples are detailed in Table \ref{ta:subsample}. 
The {\it first column} lists the selection criteria to build each subsample, 
where {\it trough min. velocity} is the minimum velocity of each trough, 
and the bottom group is parameterized using the absolute $i-$band magnitude 
at $z=2$ of the quasars, as reported in DR12Q \citep{Paris2016}. 
The {\it second column} in Table \ref{ta:subsample}, quotes 
the mean value for each selection parameter, and the {\it third one} the 
values of the intercept used for the calculation of the $a$ parameters in the 
absorption profile (Eq.~\ref{eq:fit} and \S~\ref{sec:parameter}). 
The {\it fourth column} denotes the number 
of spectra considered for the composite of each subsample. 
We publicly release our composite spectra at \url{https://github.com/lluism/BALs}, and 
their analysis is performed in our 
companion paper \cite{Masribas2019}.

\subsection{Line Profile fitting}\label{sec:eqw}

   We describe the wavelength ranges considered for the modeling of the absorption lines and 
the continua in \S~\ref{sec:continua}, the formula for the absorption profile fitting in \S~\ref{sec:fitting}, 
and the computation of the absorption width in \S~\ref{sec:parameter}.
For the modeling of the absorption lines, we consider the set of 
atomic transitions listed in Table \ref{ta:lines}, although not all of them are detected in all 
our stacks.

\subsubsection{Line Windows and Local Continua}\label{sec:continua}

   The first step toward obtaining a reliable fit to the absorption lines  
is to renormalize the outflow composite spectra. 
This modification is required because the impact of 
neighboring absorption features and/or the Lyman-alpha forest, in some cases, results in  
absorption-free spectral regions that deviate significantly from the unity 
transmission value in the final spectra. We renormalize each 
outflow composite spectrum by dividing it by a pseudo-continuum. This is computed 
via smoothing the composites with a Gaussian kernel of width $\sigma_{\rm G}=10$ 
pixels, showing no substantial differences for other values within the range 
$5 \le \sigma_{\rm G} \le 15$. For the smoothing, we disregard the wavelength 
ranges $- 1.7 \leq (\lambda-\lambda_{\rm c})/{\rm \AA} \leq 1.0$ 
around the absorption lines, where $\lambda_{\rm c}$ denotes the line center. 
The ranges are asymmetrically centered around the lines to account for the 
presence of the line-locked component in the low-wavelength side of 
each absorption feature.  We note that this renormalization washes out 
the broad absorption wings that sometimes extend blueward of the strongest absorption transitions 
(most notable for C{\sc iv}), and thus retains mostly the central and deep regions, where the 
absorption lines are well defined. In turn, the equivalent widths of some of the strongest 
lines may be underestimated, and should not be adopted without further considerations. 
Despite this re-normalization, the final absorption spectrum is still not completely flat around the 
absorption lines in some cases. For this reason, we will use a local continuum over the absorption 
lines for the profile fitting, as described below.

The absorption profiles are modeled in a wavelength window with a half width of 4 
${\rm \AA}$ on each side of the line center, plus an additional extent of 
$500\,{\rm km\,s^{-1}}$ on the low-wavelength limit to account for the presence 
of the line-locking component. This window size is large 
enough to cover the lines and their line-locked components in all our composite 
spectra. When two or more line windows overlap, we consider a single window 
spanning from the lowest to the highest limit of the intervening windows, and 
the profiles of all the lines within this range are modeled at the 
same time. This approach results in the simultaneous modeling and measurement 
of up to five absorption lines in general. In two cases we obtain a window containing ten 
lines, in the wavelength range around $\sim 1125$ \AA$\,$, and in the one around $\sim 
1195$ \AA. We simply divide both ranges in two parts, containing five lines each, and treat 
them as separate features. Most lines in these ranges are weak, and their precise 
measurement does not impact our findings, so we do not attempt more complex 
computations.

The local continuum over each absorption feature is computed by fitting a linear regression to the flux pixels next to the absorption window. We consider the outflow transmission pixels beyond the limits 
of the line 
(lines) window, up to a distance of 9 {\rm \AA} on both sides from the line center (or the minimum and maximum line centers for the case of multiple lines). 
We require a minimum total number of 15 pixels in these ranges. This value is chosen to 
reduce the impact of spurious flux pixels while representing the regions near the 
absorption lines. If the total number of pixels in the two ranges is smaller, due to 
pixels not being considered because they belong to neighboring line windows or are affected by 
skylines, both ranges are recursively enlarged by $0.4$ {\rm \AA} until the minimum number 
is reached.

\subsubsection{Profile Estimator}\label{sec:fitting}

   The modeling of the absorption lines considering a simple feature clearly does not reproduce 
the observed profiles. The presence of the potential line-locking component in our outflow spectra 
requires, for every absorption line, the joint fit of two features: the line itself and the line-locked 
counterpart. For clarity, we will refer to these two components as {\it line} and {\it line-locking} from 
here onward, and we will account for the two contributions in all our measurements. 

     We fix the center of the line-locking considering 
that $v_{\rm c} - v_{\rm ll}\equiv \Delta_v = - 497\;{\rm km\,s^{-1}}$, where 
$v_{\rm c}$ and $v_{\rm ll}$ are the positions of the line and line-locking in velocity 
space, respectively, and the value corresponds to the separation between the two lines of 
the C{\sc iv}${\,\lambda\lambda 1548,1550}$ doublet. We choose this value because the 
stacking of C{\sc iv} troughs results in the enhancement of the C{\sc iv} line-locking 
feature, while it dissipates the contribution from other doublets.
We have tested that variations of $\le 15-20\;{\rm km\,s^{-1}}$ around the nominal C{\sc iv} 
separation value (corresponding to a $\sim 22-29 \,\%$ of the pixel size) generally do not 
produce significant differences in the fits.  We also require that the depth ratio between the line 
and line-locking components (i.e. $d/b$ below) remain the same when fitting lines belonging 
to the same atomic doublet.

      We perform a least-squares fit to each line and line-locking pair assuming a profile of the form
\begin{align}\label{eq:fit}
F_\lambda = C_\lambda\, {\rm exp} & \left[  -b\, \exp{\frac{-(\lambda-\lambda_{\rm c})^2}{2a^2}} \right. 
\\ \nonumber
& ~ \left.  -d\, \exp{\frac{-(\lambda-\lambda_{\rm ll})^2}{2a^2}} \right] ~ ,
\end{align}
where the position of the line-locking in wavelength space is obtained with the relation  
$\lambda_{\rm ll} = \lambda_{\rm c} \,(1+\Delta_v/c)$, noting the negative sign of $\Delta_v$. 
The parameter $C_\lambda$ in Eq~\ref{eq:fit} represents the local continuum at a wavelength 
$\lambda$ in the renormalized rest-frame outflow composite spectrum, and $b$, $d$ and $a$ are free 
parameters. The $a$ parameter is set to be the same for the line and line-locking, which yields a 
substantial improvement of the fits, especially when multiple features are fitted at the same time and 
degeneracies between them can occur. Furthermore, we do not observe significant differences 
between the paired $a$ parameters when they are allowed to be different. However, the $a$ 
parameter is related to the width of the absorption features, which is mostly driven by the 
spectrograph resolution in the BOSS spectra, except for the broadest features. Because of this 
dependence on resolution, differences between the widths of line and line-locking, not detectable 
in our spectra, can exist. It is important that future (higher-resolution) analyses 
revisit these values as the line width carries relevant information on the kinematic properties of the 
media.

\subsubsection{Fixing the Line Widths}\label{sec:parameter}

    In practice, we fix the value of the parameter $a$ in our calculations, taking into account its 
dependence on the spectrograph resolution mentioned above and the evolution of the latter with 
wavelength. In other words, instead of using the $a$ parameters resulting from the profile fit that 
show a large dispersion, we will use these fitted values to build a relation with wavelength that will 
yield the value of  $a$ at any wavelength. 
This approach yields a reduction in the number of free parameters and, in turn, an improvement of 
the fits to the absorption profiles because the model is less sensitive to the effect of noise and other 
contaminants.

We first parametrize the evolution with wavelength in 34 of our outflow composite spectra to cover a broad range of trough parameter values and to assess possible dependences of the evolution on these parameters. We consider the total sample, the six spectra for the subsamples in outflow redshift, 
the six for the trough width,  four for the low velocity limit of troughs, the six for quasar magnitudes, and the eleven for the outflow velocities, all of them detailed in Table \ref{ta:subsample}. 
We fit all the features in Table \ref{ta:lines} for each of these spectra, with the $a$ parameter for every line and line-locking pair allowed to be free, and keep the $a$ values when the fit exists (we 
consider that a fit does not exist and ignore its outcome when it returns unphysical, i.e., null or negative, values for some of the parameters). An increase of the $a$ 
parameter with wavelength is apparent in all cases. However, a large scatter in the individual $a$ values,  partially driven by degeneracies between blended absorbers and by high noise in the spectra of some stacks that especially affects the weakest features, does not enable us to constrain the wavelength dependence with accuracy in the individual stacks. Thus, we decide to parameterize the slope of the 
evolution with wavelength by using the mean of the $a$ values from the 34 subsamples, which is consistent with the lack of visible differences in the evolutions between subsamples or between low- and high-ionization species. We finally fit a linear regression to these mean values considering the still large error bars 
arising from the dispersion of values around the mean, and use its slope of $m=3.4\times 10^{-4}$ for the computation of the $a$ parameter in all the composite subsamples. This slope is similar to the one we obtained for DLAs in \citealt{Masribas2016c} ($2.3\times 10^{-4}$), indicating that the evolution is mostly driven by the values of the wavelength and not by  physical effects specific of outflow 
environments.

\begin{figure}\center 
\includegraphics[width=0.48\textwidth]{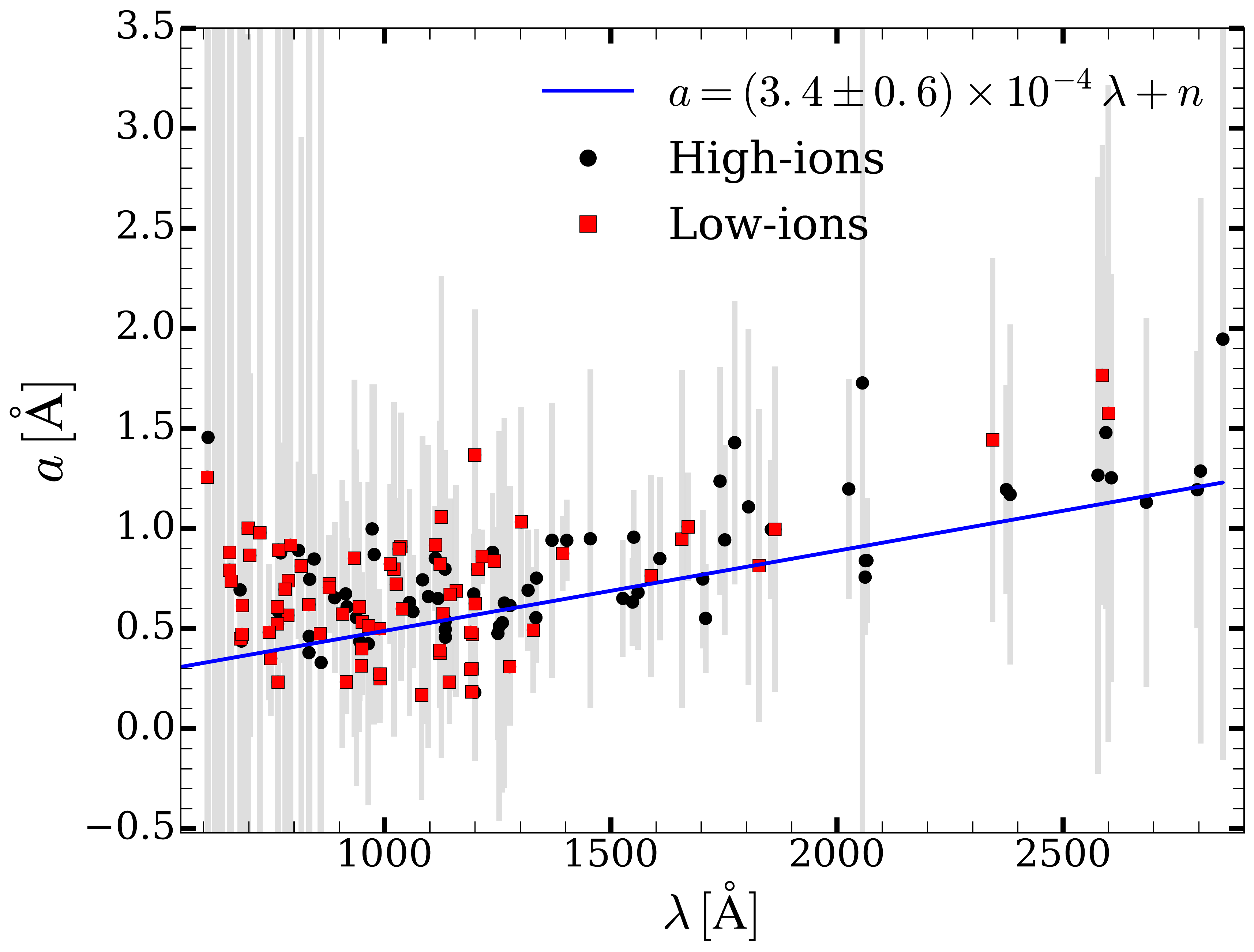}
\caption{ Mean values of the $a$ parameters with wavelength, computed from the 34 
composite outflow spectra detailed in the main text. There are no apparent differences 
between the trends for low- ({\it red squares}) and high-ionization ({\it black dots}) 
species, and a general increase of $a$ with wavelength is visible although with a large 
scatter ({\it gray lines}). The {\it blue line} denotes the linear fit that yields the slope 
value $m=3.4\times 10^{-4}$, used to obtain the $a$ parameter in the individual stacks.}
\label{fig:aparam}
\end{figure}

   Figure \ref{fig:aparam} displays the mean value of the $a$ parameters with wavelength, computed using the 34 composite spectra of the subsamples described above. 
A general increase of $a$ with wavelength is visible although with significant scatter (error bars denoted by {\it gray lines}), and there are no apparent differences between low- ({\it red squares}) and 
high-ionization ({\it black dots}) species. The {\it blue line} denotes the linear fit to the data, considering  the uncertainty, that yields the slope value of $m=3.4\times 10^{-4}$.

With the value of the slope now fixed, we again apply a linear regression to the $a$ values of 
each subsample to obtain the individual best-fit value of the intercept, $n$.
In this case, we only consider the $a$ parameters of the strong transitions H{\sc i}$\,\lambda1215$, N{\sc v}$\,\lambda1238$, N{\sc v}$\,\lambda1242$, Si{\sc iv}$\,\lambda1393$, Si{\sc iv}$\,\lambda1402$, and C{\sc iv}$\,\lambda1548$, where the $a$ parameters are measured with the highest precision. In all cases, we visually inspect the fits to these absorption features and discard 
those that appear dubious or affected by contaminants. The best-fit values of the $n$ parameters for each subsample are reported in Table \ref{ta:subsample}. Finally, the value of the $a$ parameter for an absorption feature at wavelength $\lambda$ can be obtained as 
\begin{equation}\label{eq:aeq}
a  = (3.4 \pm 0.6)\times 10^{-4}\, \lambda   + n ~,
\end{equation}
where $n$ takes on the values in Table \ref{ta:subsample} for each spectrum, and the parameters are in units of ${\rm \AA}$ngstrom.

\begin{figure*}\center 
\includegraphics[width=1.\textwidth]{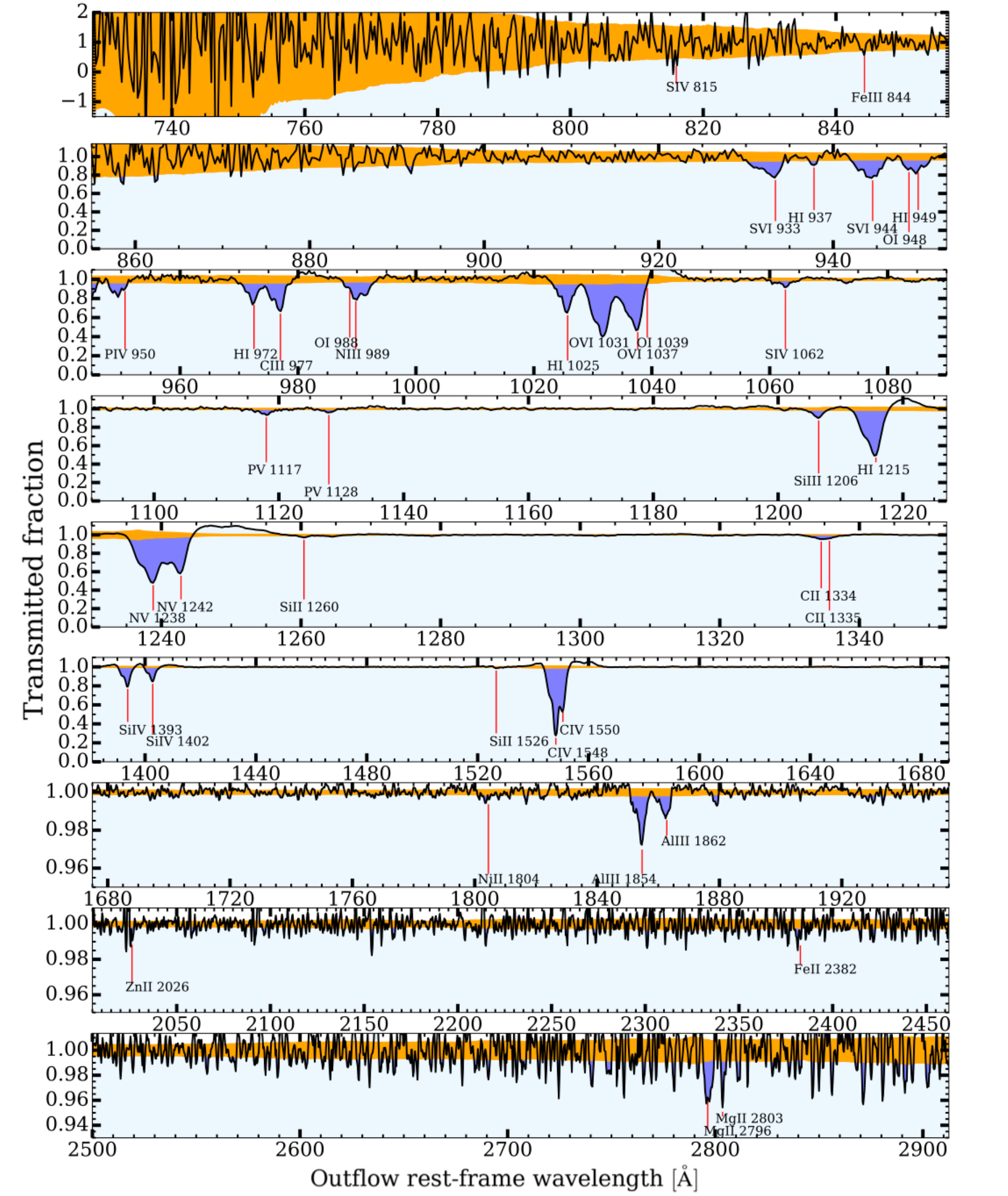}
\caption{Outflow composite spectrum for the subsample with outflow velocities in the range 
$1\,500 \leq v  < 3\,000$ ${\rm km\,s^{-1}}$. The {\it light blue} color represents 
the transmitted quasar flux and the {\it dark blue} the absorbed regions with 
transmission below the unity value. The absorption lines beyond the one-sigma noise ({\it orange 
band}) are labeled 
for identification, and the axes in each range are independently scaled for 
visualization. The line-locking component is visible in the low-wavelength side of the 
strongest absorption features. This spectrum was normalized 
for visualization by using a simple smooth continuum on the original spectrum.  
A flux excess next to some of the strongest absorption features arises from the differences 
between the mean quasar spectrum used for the normalization of the composites 
and the individual quasar spectra.}
\label{fig:dissect}
\end{figure*}

\section{Dissection of an Outflow Composite Spectrum}\label{sec:dissection}

  Figure \ref{fig:dissect} displays the outflow composite spectrum resulting 
from the subsample of troughs with outflow velocities within the range 
$1\,500 \leq v  < 3\,000$ ${\rm km\,s^{-1}}$. The {\it light blue} color 
represents the transmitted flux regions, and the {\it dark blue} highlights the 
absorbed ones where the transmission falls below the $1\sigma$ noise shown in 
{\it orange}. This 
noise is computed for each pixel, as the median of the variance of the flux around the unity 
level, and considering a region centered at the pixel with a half-window size of $25$ \AA. 
The pixels in regions corresponding to absorption lines (\S~\ref{sec:continua}) are discarded 
for the noise calculation. Note that the scales of the axes for each range are set independently  
for visualization purposes. 

  This spectrum was renormalized 
using a pseudo-continuum as described in \S~\ref{sec:continua}, but the local 
continua around the absorption features used for the calculation of the equivalent 
widths were not applied because they depend on each feature. Regions with transmission 
above the unity, visible on the right-hand side of the absorption features of O{\sc i}$\,\lambda 
1039$, H{\sc i}$\,\lambda 1215$, N{\sc v}$\,\lambda 1242$ and C{\sc iv}$\,\lambda 
1550$, appear due to differences between the mean quasar spectrum used for the normalization 
of the composites and the individual quasar spectra. In general, this  
flux excess indicates that the mean quasar spectrum presents weaker emission lines than 
those in the individual quasars. However, the impact of weighting the spectra also causes this 
effect, because the higher S/N is typically related to the brighter quasars which, in turn, have 
weaker emission lines due to the Baldwin effect \citep[][see Figure 3 in \citealt{Masribas2016c} 
and the associated text for details]{Baldwin1977}. 
We do not apply further corrections because this effect does not impact our results concerning the line 
locking, and the equivalent widths of the lines are well fitted given the fix value of their widths. 

   We have labeled the strong atomic transitions from Table \ref{ta:lines}, 
a number of them identifiable deep into the Lyman-alpha forest region. 
The strongest features are the well-resolved doublets of the high-ionization 
species S{\sc vi}$\,\lambda\lambda933,944$, O{\sc vi}$\,\lambda\lambda1031,1037$, 
N{\sc v}$\,\lambda\lambda1238,1242$ and  C{\sc iv}$\,\lambda\lambda1548,1550$. These ions 
represent the typical effect from the strong radiation field affecting the outflow in general. However, 
other transitions of high- and low-ionization species are clearly visible, 
such as the first five hydrogen Lyman series lines from Lyman-alpha down to 
Lyman-epsilon, i.e., H{\sc i}$\,\lambda 937$, and 
C{\sc iii}$\,\lambda 977$, N{\sc iii}$\,\lambda 989$, Si{\sc iii}$\,\lambda 1206$, 
and the doublets P{\sc v}$\,\lambda\lambda1117,1128$, Si{\sc iv}$\,\lambda\lambda1393,1402$, 
Al{\sc iii}$\,\lambda\lambda1854,1862$ and Mg{\sc ii}$\,\lambda\lambda2796,2803$. 
We have also labeled transitions  of the 
low-ionization species Ni{\sc ii}, Fe{\sc ii}, O{\sc i}, Si{\sc ii}, C{\sc ii}. 
We find no evidence for the presence of molecular ${\rm H_2}$ features, 
although the line-locking component on the left-hand side of the absorption lines 
may overlap and mask the molecular hydrogen signatures. 
We study the physical properties of the outflows, by analyzing this and the other composite 
spectra, in our companion paper \cite{Masribas2019}. 

     An absorption feature on the 
low-wavelength side of all lines, and blended with them,  is clearly visible for the case of 
the strongest transitions in Figure \ref{fig:dissect}. We identify this feature as the C{\sc iv} line-locking 
component of the corresponding lines, and analyze it in more detail below.

 \begin{figure*}\centering 
\includegraphics[width=1\textwidth]{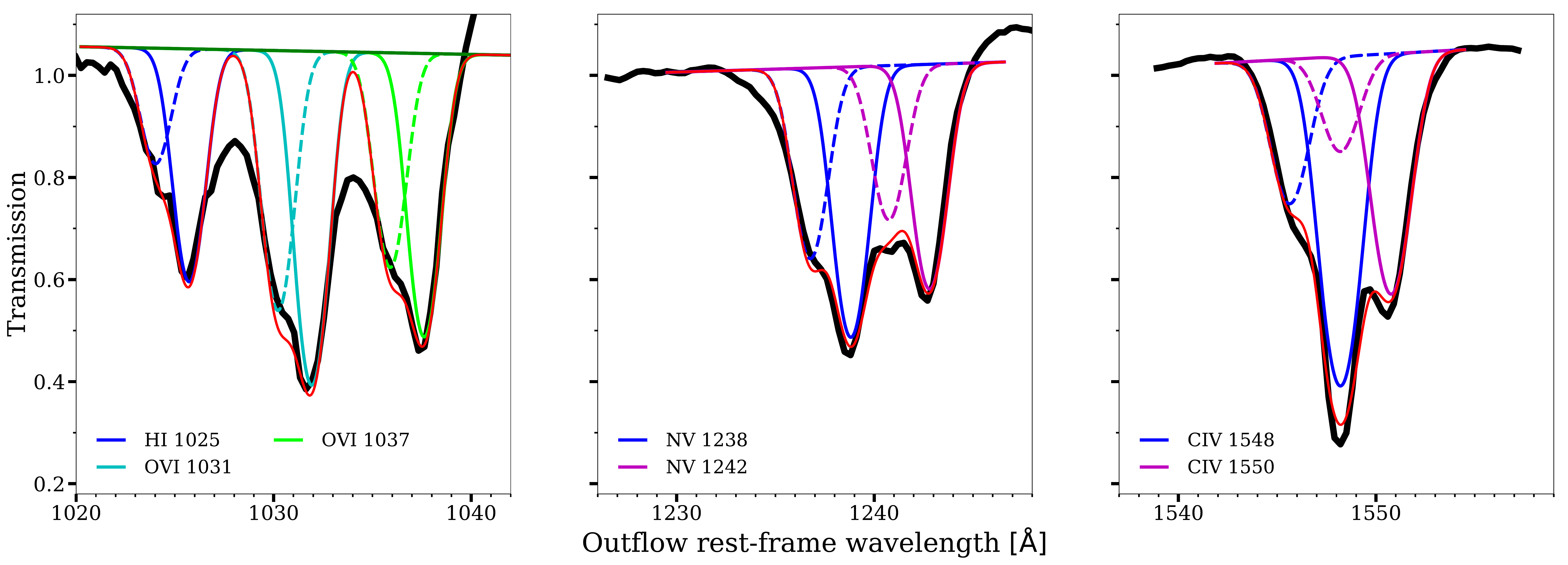}
\caption{Fits to the strongest absorption features in the composite spectrum of the 
subsample with outflow velocity $1\,500 \leq v  < 3\,000$ ${\rm km\,s^{-1}}$ 
(spectrum in Figure \ref{fig:dissect}). The {\it colored solid curves} denote the fit to 
the individual lines, and the {\it dashed lines} are the fits to the corresponding 
line-lockings, following the same color code. The {\it red curves} represent 
the total fit in each panel. The line-locking components are visibly required to obtain 
good fits.  The leftmost panel shows regions between the lines that are not covered 
by the fit. This is due to the presence of a number of weak low-ionization lines in 
these regions, mostly from O{\sc i} and C{\sc ii}, not included in our simple approach (see main text).}
\label{fig:fitlock}
\end{figure*}

 \begin{figure*}\centering 
\includegraphics[width=1\textwidth]{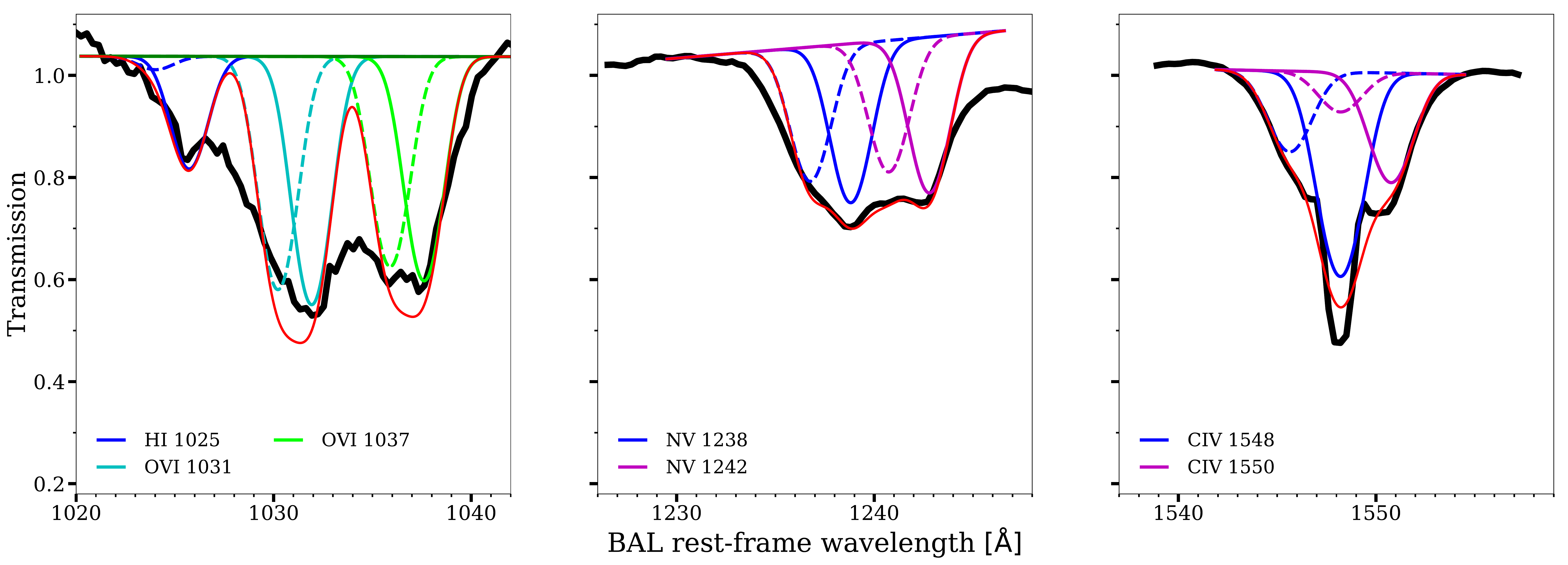}
\caption{Same as Figure \ref{fig:fitlock} but for a sample constructed only with BAL 
troughs, i.e., absorption troughs blueshifted by $ > 3\,000\,{\rm km\,s^{-1}}$ from the respective 
quasar emission line, and with a full width at half {\it minimum} $\gtrsim 2\,000\,{\rm km\,s^{-1}}$. }
\label{fig:balfit}
\end{figure*}

\section{Line Locking}\label{sec:locking}

  We analyze the characteristics and nature of the line-locking components below. In 
\S~\ref{sec:civlock}, we discuss the C{\sc iv}-driven features visible in our spectra, 
and search for signatures of {\rm O}{\sc vi} and {\rm N}{\sc v} in \S~\ref{sec:nvlock}.

\subsection{{\rm C}{\sc iv} Line Locking}\label{sec:civlock}

  Figure \ref{fig:fitlock} illustrates the fits to three of the strongest 
absorption features in the spectrum of Figure \ref{fig:dissect}. From left to right, 
the panels cover the approximate wavelength ranges $1\,020-1\,040\,{\rm \AA}$, 
$1\,230-1\,250\,{\rm \AA}$, and $1\,540-1\,560\,{\rm \AA}$. The 
{\it colored solid curves} denote the fit to the individual lines,   
and the {\it dashed curves} are the fits to the corresponding 
line-lockings, following the same color code. The {\it red curves} represent 
the total fit in each panel, and the local continua are visible as 
{\it straight lines} over the features, where the colors simply 
result from the superposition of the individual curves and have no meaning. 
The C{\sc iv} line-locking components at $497\,{\rm km\,s^{-1}}$ from the corresponding 
lines are clearly visible. This occurs for any absorption feature  
visible beyond the noise in all our composite spectra, regardless of 
the species or the ionization state, as expected from the theory of radiative 
acceleration in quasar outflows \citep[e.g.,][]{Arav1994,Arav1994b}. 

    The leftmost panel in Figure \ref{fig:fitlock} shows regions between the 
absorption lines that are not fitted by our model. This is because we have selected the 
strongest lines in this region to illustrate the presence of line-locking, but we have not considered 
a number of weaker transitions that inhabit the regions in between, and that sometimes overlap 
with the strongest lines. Our simple methodology does not yield reliable fits in cases of many strongly 
blended lines. The two strongest omitted transitions are C{\sc ii}$\,\lambda 1036$ and the excited-state 
line C{\sc ii}$\,\lambda 1037$. However, the effect of these two lines together should be small,  
similar to the signal observed for C{\sc ii}$\,\lambda 1334$ and C{\sc ii}$\,\lambda 1335$ in Figure 
\ref{fig:dissect} because the oscillator strengths for all these lines are almost the same 
(Table \ref{ta:lines}). Other O{\sc i} transitions also impact those regions, with strengths about 
half of that of O{\sc i}$\,\lambda 1302$ in Figure \ref{fig:dissect}, according to the relation of 
oscillator strengths. The three strongest lines that we consider for the fit are in many of the 36 
stacks enough to produce a complete fit (see Figures in \S~\ref{sec:fitapp} in the Appendix). 

    In cases of blending or weak absorption features, the fit is sometimes 
better without the line-locking (or line), but we attribute this result to 
degeneracies between overlapping transitions or to the effect of noise, 
respectively, and not to physical causes. Some equivalent widths can be contaminated by 
these effects and yield unreliable measurements. Relevant (strong) transitions 
potentially affected by degeneracies are H{\sc i}$\,\lambda 949$,  H{\sc i}$\,\lambda 1025$, 
O{\sc vi}$\,\lambda 1037$,  and the 
C{\sc iv}$\,\lambda\lambda 1548,1550$ doublet. Instead, the Lyman-alpha line, 
i.e., H{\sc i}$\,\lambda 1215$, and the doublets N{\sc v}$\,\lambda\lambda1238,1242$ 
and Si{\sc iv}$\,\lambda\lambda1393,1402,$ are well-measured in most of our  
composite spectra. 

   The line-depth ratios between the line-locking and line features  of Figure \ref{fig:fitlock} 
vary broadly between $\sim 0.3$ and $\sim 0.8$, and the scatter increases when considering 
more lines. This result suggests that line locking occurs in a fraction of quasar outflows, the 
exact number  depending on the uncertain value of the ratio, which can also be (partially) driven 
by differences between the optical depths of the line and the line locking. From the composite 
spectra of about $19\,500$ narrow  ($<200\,{\rm km\,s^{-1}}$) C{\sc iv} absorbers in SDSS 
quasars showing BALs, \cite{Bowler2014} inferred that $50-75\,\%$ of C{\sc iv} systems in  
outflows are line locked, which broadly agrees with our ratio values. 

    We show in \S~\ref{sec:fitapp} in the Appendix the fits to the previous three wavelength 
ranges, as well as that covering the \lya absorption line and line-locking, for all our composites. 
The line-locking component is distinguishable in most cases. There are some cases where a  
good fit cannot be obtained due to the effects of noise or contaminants. These cases are 
left as blanks in the plots. 

\subsubsection{Line Locking in BALs}\label{sec:ballock}

    Our work aims at analyzing the quasar outflow properties based on the physical 
properties of the individual C{\sc iv} absorption systems, and not on the nomenclature often 
assigned to the troughs themselves (e.g., BALs, mini-BALs, or NALs). We do so because these 
names may all refer to the same type of quasar outflows although accounting for different 
properties \citep[][see also how these different terms can be accounted for considering the 
differences in outflow velocity and trough width in \citealt{Masribas2019}]{Hamann2018}. 
 Due to the extended use of some of these terms, however, we briefly show here the presence 
of line-locking in an additional sample built from the 
strict definition of broad absorption line (BAL) systems. Specifically, BALs refer to absorption troughs 
blueshifted by $ > 3\,000\,{\rm km\,s^{-1}}$ from the respective quasar emission line, and 
with a full width at half {\it minimum} of $\gtrsim 2\,000\,{\rm km\,s^{-1}}$ \citep{Weymann1991}. 

     Figure \ref{fig:balfit} shows the fits to the same absorption features as in Figure \ref{fig:fitlock} 
but for the case of BALs. The presence of the line-locking components is again required to obtain 
good fits to the overall absorption features.

\subsection{{\rm O}{\sc vi} and {\rm N}{\sc v} Line Locking}\label{sec:nvlock}

   We search here for line-locking signatures of the O{\sc vi} and N{\sc v} 
doublets, which may have been washed out in our C{\sc iv}-trough composite spectra. 
These ions may potentially participate in the radiative acceleration process  
through line locking, given their high cosmic abundances and the similar strength 
of their absorption features compared to that of C{\sc iv} (see, e.g., Figure 
\ref{fig:fitlock} or the strong effects by these ions already suggested on the 
quasar spectrum of Figure \ref{fig:cont}). This argument is also 
supported by the results by \cite{Bowler2014}, who modeled the photoionization 
state of quasar outflows and found that the optical depths for O{\sc vi} and 
N{\sc v} are similar to that of C{\sc iv} (their Figure 12). Furthermore, 
an additional line optical-depth autocorrelation analysis by the same authors 
confirmed the importance of these two ions to the overall opacity, although no 
absorption signatures were detected in their work.

\begin{figure}\center 
\includegraphics[width=0.48\textwidth]{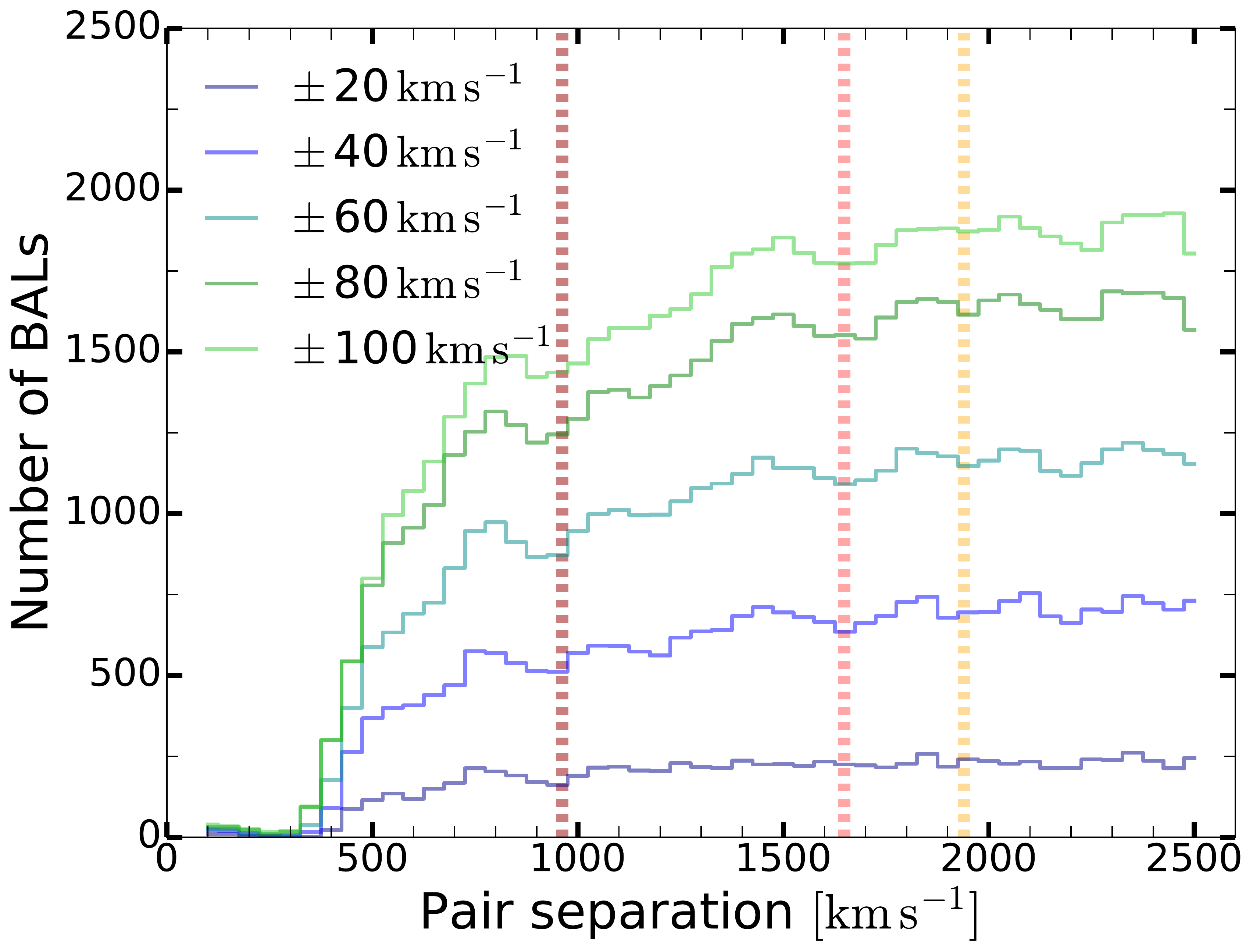}
\caption{Number of pairs of C{\sc iv} troughs separated by a given distance, from 
100 to $2\,500$ ${\rm km\,s^{-1}}$, in steps of 50 ${\rm km\,s^{-1}}$. We enable 
five offset values from the nominal positions of the troughs. The vertical dashed 
lines indicate the distances corresponding to the separations of the N{\sc v} (962 
${\rm km\,s^{-1}}$), O{\sc vi} ($1\,648$ ${\rm km\,s^{-1}}$), and Si{\sc iv} ($1\,940$ 
${\rm km\,s^{-1}}$) doublets. No excess in the number of troughs is detected at the line-locking 
separations.}
\label{fig:fitovi}
\end{figure}

   As our absorber catalog is only for C{\sc iv} systems and not for other species, we cannot 
perform the same analysis to search for line-locking of other elements. Instead, we compute the 
number of pairs of C{\sc iv}  troughs in 
our catalog that are separated by a distance from 100 to $2\,500$ ${\rm km\,s^{-1}}$ in steps of 
50 ${\rm km\,s^{-1}}$. If line locking from other species occurs with frequency, and its signature is 
strong, the number of troughs at the locking separation should be  larger than at other 
distances. Figure \ref{fig:fitovi} shows the results of this calculation, where we have allowed for 
different offset values from the exact position of the expected troughs. The dashed vertical lines 
denote the separations of the N{\sc v}, O{\sc vi}, and Si{\sc iv} doublets. No excess in the number 
of troughs at the line-locking separations is detected in any case.  

    We  conclude from this analysis that the N{\sc v}, Si{\sc iv} and O{\sc vi} line-locking 
effects, if present, are not frequent and/or strong enough to produce an absorption feature 
detectable in the BOSS spectra. 

\section{Discussion}\label{sec:discussion}

  We discuss the reality of the C{\sc iv} line-locking features in \S~\ref{sec:really},  
the possible causes for the non-detection of line locking from other species in 
\S~\ref{sec:nvovino}, and compare with previous work in \S~\ref{sec:comp}.

\begin{figure*}\center 
\includegraphics[width=1.\textwidth]{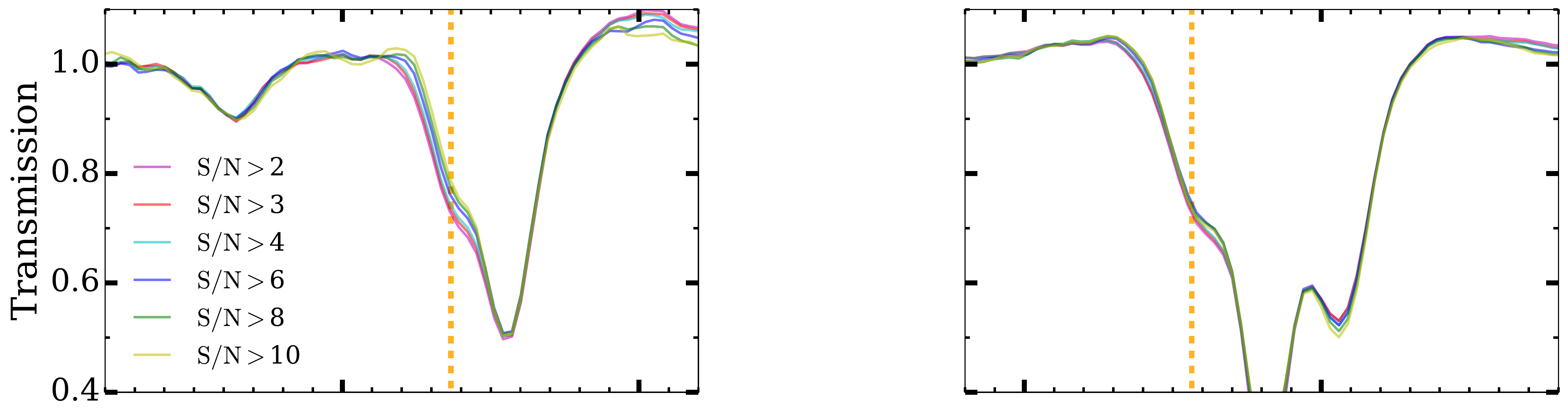}
\includegraphics[width=1.\textwidth]{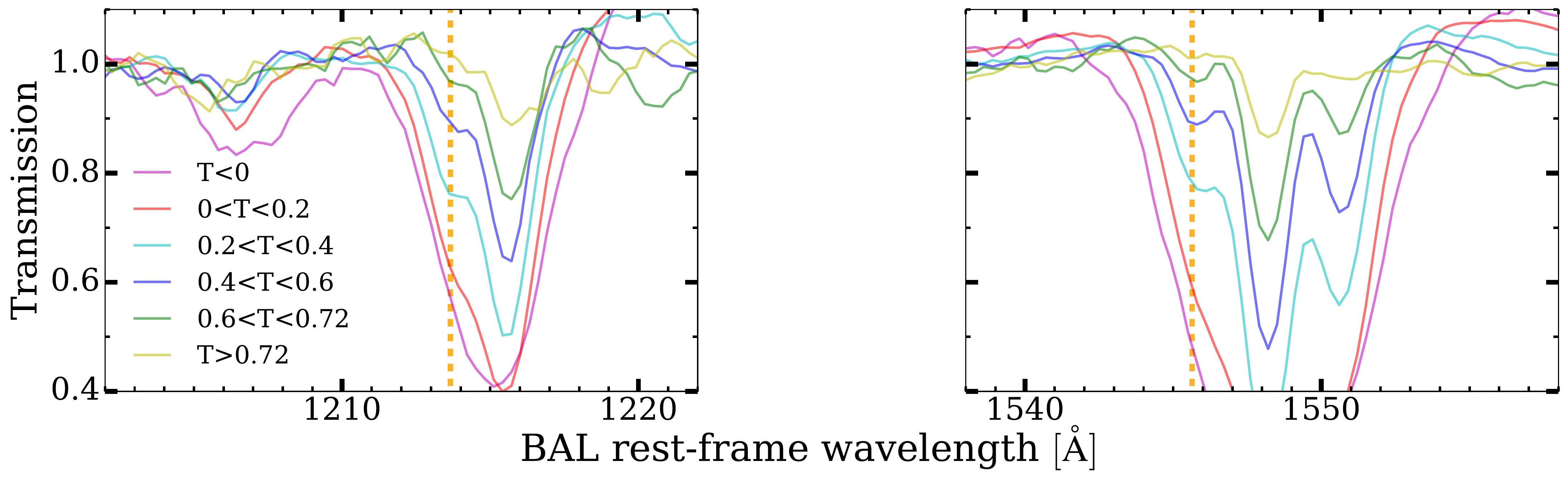}
\caption{Comparison 
of composite spectra drawn from the trough subsample with outflow velocity $1\,500 \leq v  
< 3\,000$ ${\rm km\,s^{-1}}$ and considering different S/N thresholds (given for the 
whole spectrum in DR12Q; {\it upper panel}), and minimum-flux values (transmission) 
in the absorption troughs (values from the DR12QBAL; {\it lower panel}). The {\it left} 
and {\it right panels} illustrate the regions blueward the rest-frame absorption 
Ly$\alpha$ and C{\sc iv} doublet features, and the {\it dashed lines} denote the position 
of the C{\sc iv}-locked components. The line-locking feature is visible in all cases 
which suggests that it is real and not an artifact from the combination of doublets 
centered at different lines of the C{\sc iv} doublet due to the noise or line strength.}
\label{fig:locktest}
\end{figure*}

\subsection{Is It Really Line Locking?}\label{sec:really}

   The presence of a C{\sc iv} line-locking-like feature on the blue 
side of almost all the absorption lines is apparent in our stacks, but we found no 
conclusive evidence for the existence of the O{\sc vi}, N{\sc v} or other  
line-locking effects. We revisit here the effects of our methodology that may 
compromise the reality of the C{\sc iv} line-locking.   We use the same composite 
spectrum analyzed above for consistency, the subsample with outflow velocity $1\,500 \leq v 
 < 3\,000$ ${\rm km\,s^{-1}}$. We do not compute the additional $\sim 400$ stacks that arise 
from splitting each of the 36 spectra in the subgroups detailed below for testing all the composites.  
The impact from noise can be different for each composite, but considering the figures in 
\S~\ref{sec:fitapp} in the 
Appendix, it seems difficult that noise can drive all the observations. 

   In \S~\ref{sec:stack} we checked that centering the outflow spectra at the position 
of the minimum flux within the absorption troughs yields the most refined 
(spectrally-resolved) composite spectra. This result is consistent with the findings 
by \cite{Perrotta2018}, and suggests that the origin of the minimum-flux is mostly 
physical, i.e., due to the presence of the stronger absorption component among 
the absorption trough, and not driven by noise. In the case of one single C{\sc iv} 
doublet, the blue (shorter wavelength) line is expected to be within one and two 
times stronger than the red line, the exact value set by atomic physics 
(i.e., by the ratio between the two oscillator strengths) and the level of 
saturation of the lines. In a noise-free doublet therefore, the minimum flux 
will always appear at the position of the blue line, except in the cases where 
the two lines are saturated and have the same minimum flux. 
Thus, although the minimum flux can be mostly governed by the 
lines, in cases of strong saturation these minima can be identified at the position 
of the red line of the doublet, with the corresponding location of the blue line at 
the position of the expected line-locking component. If this occurs a large number 
of times, a residual feature can be visible in the stacks. Furthermore, the effect 
of high noise levels could contribute to increase this problem. 

   We analyze the aforementioned potential impacts from the noise and the strength 
of the absorption lines below. Figure \ref{fig:locktest} shows the comparison 
of composite spectra drawn from the trough subsample with outflow velocity $1\,500 \leq v  
< 3\,000$ ${\rm km\,s^{-1}}$, considering different S/N thresholds (given for the 
whole spectrum in DR12Q; {\it upper panel}), and minimum-flux values (transmission) 
of the troughs (values from the DR12QBAL; {\it lower panel}). The {\it left} 
and {\it right panels} illustrate the regions blueward the rest-frame absorption 
Ly$\alpha$ line and the C{\sc iv} doublet, respectively, and the {\it dashed lines} 
denote the position of the C{\sc iv} line-locked components. The {\it top panels} show 
that the line-locking features remain almost constant when increasing the S/N of the 
spectra, reflecting the small impact of the noise on our stacks. The line-locked 
features are also visible for the different ranges of transmission in the {\it bottom 
panels}, indicating the almost negligible effect from strong saturation that could 
increase the fraction of minimum-flux positions derived from the red line of the 
doublets. More important, the ratios between the lines and line-lockings remain 
almost the same in all cases, highlighting the relation between the two components 
driven by the line-locking effect. 

   We conclude that the absorption features detected 
blueward of the absorption lines  in the deepest regions of the absorption features of 
our composite outflow spectra are the result of the 
C{\sc iv} line-locking effect, which demonstrates the significant contribution from 
radiation to the acceleration of quasar outflows. 

\subsection{On the Non-detection of N{\sc v} and O{\sc vi} line-locking}\label{sec:nvovino}

    It is possible that physical effects do prevent line locking from N{\sc v} and O{\sc vi} to occur.
Overionization of the medium inhabited by these species can suppress radiative acceleration 
from theses species because it results in too low opacities and optically thin media 
\citep{Murray1995,Chelouche2003,Leighly2004,proga2004,proga2007,Laor2014,Wang2015}. 
However, C{\sc iv} has a lower ionization potential, and typically traces cooler and denser 
media than the oxygen and nitrogen ions. This region could also be shielded 
(by the highest ions themselves) from the high-energy radiation field and 
thus enable radiative acceleration from C{\sc iv} line locking to take place. 
We discuss the potential impact of the outflow properties and its structure on the 
presence of the N{\sc v} and O{\sc vi} absorption features in \cite{Masribas2019}.

      The existence of O{\sc vi}, N{\sc v} and other line-locking signatures would be best assessed by 
using absorption troughs of these species directly identified in the individual quasar spectra, instead 
of using those of C{\sc iv}. The search for N{\sc v} and O{\sc vi} troughs, however, is difficult 
because they are in regions of the quasar spectrum affected by the Ly$\alpha$ emission line 
and forest, respectively. Searching for Si{\sc iv} features may be easier since a fraction of them 
will likely reside in the line-free region of the rest-frame quasar spectrum. However, the strength of this 
doublet is significantly smaller than those of O{\sc vi}, N{\sc v} and C{\sc iv} (Fig. 1),  and so its  
line-locking feature could also be weak and difficult to detect. A catalog of Si{\sc iv} absorption 
troughs recently built by \cite{Guo2019} will enable such a calculation. 
Alternatively, an autocorrelation analysis in the real BOSS spectra may reveal the characteristic 
distances where absorption troughs are located, although this requires obtaining the continuum 
for each individual quasar spectrum with precision, including the region within the 
Lyman-alpha forest where the continuum is difficult to obtain \citep{FaucherGiguere2008}.
We defer these calculations to future work.

\subsection{Comparison to Previous Work}\label{sec:comp}

    A number of works have previously performed stacks of quasar spectra containing outflows 
\citep[e.g.,][]{Baskin2013,Hamann2018}. These works have mostly focused on the analysis 
of the composite absorption troughs, while our work concerns the detection of the line-locking 
signatures. To fit the absorption lines, we have concentrated our attention to the regions where the 
lines are expected; more specifically, we have disregarded broad absorption wings that can appear on the 
blue side of some of the strongest absorption lines, and for stacks considering the broadest troughs 
(\S~\ref{sec:continua}). Therefore, our stacks can present, in some cases, shallower and narrower 
absorption lines when compared to those accounted for in previous works.

     In addition, our stacking methodology enables us to identify the different components of the 
atomic doublets in most composites, while these appear less resolved in other works. The reason 
behind this effect may arise from the fact that we align the spectra for the stacks considering the 
position of the minimum flux within the troughs directly, without smoothing the flux to account for 
the impact of noise. If this is true, this might also be the reason why line-locking has not been detected 
in most previous works. Only \cite{Bowler2014} have found the signature of line locking in stacked 
spectra. These authors stacked narrow absorbers ($<200\,{\rm km\,s^{-1}}$ C{\sc iv} troughs) in 
SDSS quasars that contain BALs. 
Although this implies that the quasars have outflows, they do not address the case of broad 
absorbers because the line and line-locking features could not be resolved. Thus, our work is the 
first one demonstrating the presence of line-locking in quasar outflows, from narrow to broad 
absorbers, and for different trough and quasar properties. It is important to mention, 
however, that \cite{Baskin2013} also reported tentative evidence for radiative acceleration, in their 
appendix B, resulting from the relation between the depth of the absorption profiles and the 
luminosity of the sources.

\section{Conclusion}\label{sec:conclusions}

   We have built 36 outflow composite spectra by stacking broad ($>450\,{\rm 
km\,s^{-1}}$) absorption line systems in the spectra of SDSS-III/BOSS DR12 quasars. 
We have computed the composites considering bins in outflow velocity, width of the troughs, 
degree of detachment between the troughs and the quasars, and quasar redshift and brightness. 
For every spectrum, we modeled the absorption profiles considering the line and line-locking 
components for a large number of atomic transitions. This atomic dataset and the 
composite outflow spectra are publicly available 
at \url{https://github.com/lluism/BALs}. 
Our results can be summarized as follows:

\begin{itemize}

\item An absorption feature is visible on the blue side of all the strong absorption lines and 
         in all our 36 composite spectra. This feature is well fitted assuming that it is the C{\sc iv} 
         line-locking component of the respective absorption lines, at a distance of 
         $497\,{\rm km\,s^{-1}}$ from them, corresponding to the C{\sc iv} doublet separation. 
         
\item The detectability of the C{\sc iv} line-locking feature compared to that of the line 
	does not  seem to depend on the signal-to-noise of the spectra or the depth of the troughs, 
	 suggesting that the locking feature is real.
	
\item We investigate the presence of line-locking features from the doublets of O{\sc vi}, 
	Si{\sc iv} and N{\sc v}, but these seem to not be present.
	
\item Our composite spectra resolve the two lines of the C{\sc iv} and other doublets, although
         the mean width value of the troughs used in the stacks is $\approx 2\,000\,{\rm km\,s^{-1}}$. 
         This implies that {\it (1)} the position of the minimum flux in the absorption troughs, which 
         we use for centering our stacks, has a physical meaning (likely representing the blue line 
         of the C{\sc iv} doublet), and {\it (2)} the broad absorption troughs may consist in the 
         superposition of narrow absorbers. 
         
\end{itemize}

   Our results indicate that radiative acceleration is a common mechanism  intervening 
in quasar outflows, and that its presence depends weakly on the characteristics 
of the winds. Models and simulations assessing the connection between the outflows 
and AGN feedback should, therefore, incorporate radiative transfer 
processes that account for the coupling between radiation and the gas, especially the resonant 
scattering of photons. The omnipresence of line locking in different wind conditions,  
also suggests that radiative acceleration might be a dominant effect in other environments. 
These could be the outflows from gamma-ray bursts \citep{Castro2010},   
quasar jets \citep{Petrucci2017}, or Seyfert galaxies, where line locking has 
been suggested and tentative detections have been reported. 
Finally, the fact that line locking is observed for the C{\sc iv} doublet but  not for other species 
may be connected to the physical properties of a characteristic multiphase structure in the outflow, 
where different species trace different regions. We present and test this hypothesis in more 
detail in a companion paper, \cite{Masribas2019}.

\section*{acknowledgements}

The initial inspiration for this work grew out of a stimulating discussion with Paul 
Martini during a visit supported by the Visitor Program at the Ohio State Center
for Cosmology and Astroparticle Physics. We are grateful to him for valuable 
ideas and comments on our paper, and to the CCAP for kind hospitality. We thank 
the anonymous referee for a detailed revision and a constructive report that 
helped improving our work. We thank Stan Owocki for sharing with us his thoughts 
and inspiring notes on the dynamics of CIV line locking.
We also thank Ainar Drews, Mattia Mina, Robert Wissing and H{\aa}vard T. Ihle for 
discussions on line-locking and statistical aspects, and Tzu-Ching Chang, 
Olivier Dor\'e, Phil Berger, Sterl Phinney, Lee Armus, Ski Antonucci, 
Brice M\'enard, Jordi Miralda Escud\'e, Sijing Shen, Johan Fynbo, Joop Schaye, 
Jason X. Prochaska, Bill Forman and Christine Jones for 
enriching conversations and suggestions. We are thankful to 
Signe Riemer-S\o rensen for many useful discussions on quasar continua. L.M.R. 
is grateful to the UCSB/MPIA ENIGMA group for their kind hospitality and, 
together with other colleagues at JPL and Caltech, for many inspiring 
discussions during this work.
This research was partially carried out at the Jet Propulsion Laboratory, 
California Institute of Technology, under a contract with the National Aeronautics 
and Space Administration.

   Funding for SDSS-III has been provided by the Alfred P. Sloan Foundation, 
the Participating Institutions, the National Science Foundation, and the U.S. 
Department of Energy Office of Science. The SDSS-III web site is 
\url{http://www.sdss3.org/}. SDSS-III is managed by the Astrophysical Research
Consortium for the Participating Institutions of the SDSS-III Collaboration 
including the University of Arizona, the Brazilian Participation Group, Brookhaven 
National Laboratory, Carnegie Mellon University, University of Florida, the French 
Participation Group, the German Participation Group, Harvard University, the 
Instituto de Astrofisica de Canarias, the Michigan State/Notre Dame/JINA 
Participation Group, Johns Hopkins University, Lawrence Berkeley National 
Laboratory, Max Planck Institute for Astrophysics, Max Planck Institute for 
Extraterrestrial Physics, New Mexico State University, New York University, 
Ohio State University, Pennsylvania State University, University of Portsmouth, 
Princeton University, the Spanish Participation Group, University of Tokyo, 
University of Utah, Vanderbilt University, University of Virginia,
University of Washington, and Yale University.

\bibliographystyle{apj}
\bibliography{locking}\label{References}


\appendix
\section{Fits to the Strongest Features in All the Outflow Composites}\label{sec:fitapp}

  Figure \ref{fig:fitall} illustrates the fits to the four strongest absorption features in all our 
composite spectra. Cases for which a good fit cannot be obtained are left as blank.  
The color code is the same as used in Figure \ref{fig:fitlock}.

\begin{figure}
\centering
\begin{minipage}{0.3cm}
\rotatebox{90}{\small\textcolor{black}{Transmission}}
\end{minipage}%
\begin{minipage}{\dimexpr\linewidth-2.50cm\relax}%
    \centering
  \raisebox{\dimexpr-.5\height-1em}{\includegraphics[scale=0.17]{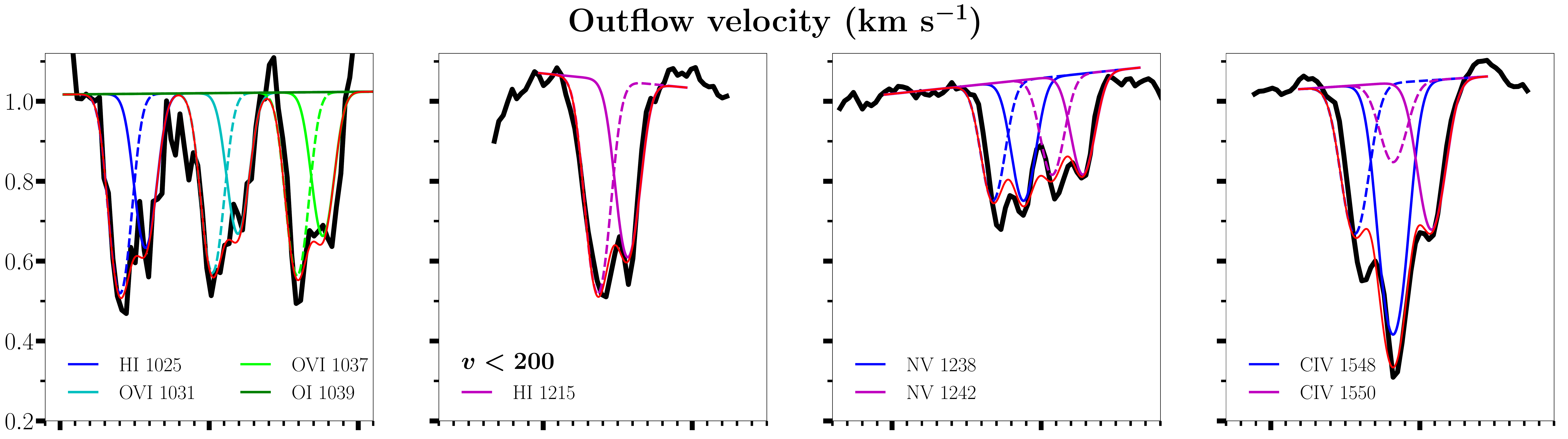}}\
  \raisebox{\dimexpr-.5\height-1em}{\includegraphics[scale=0.17]{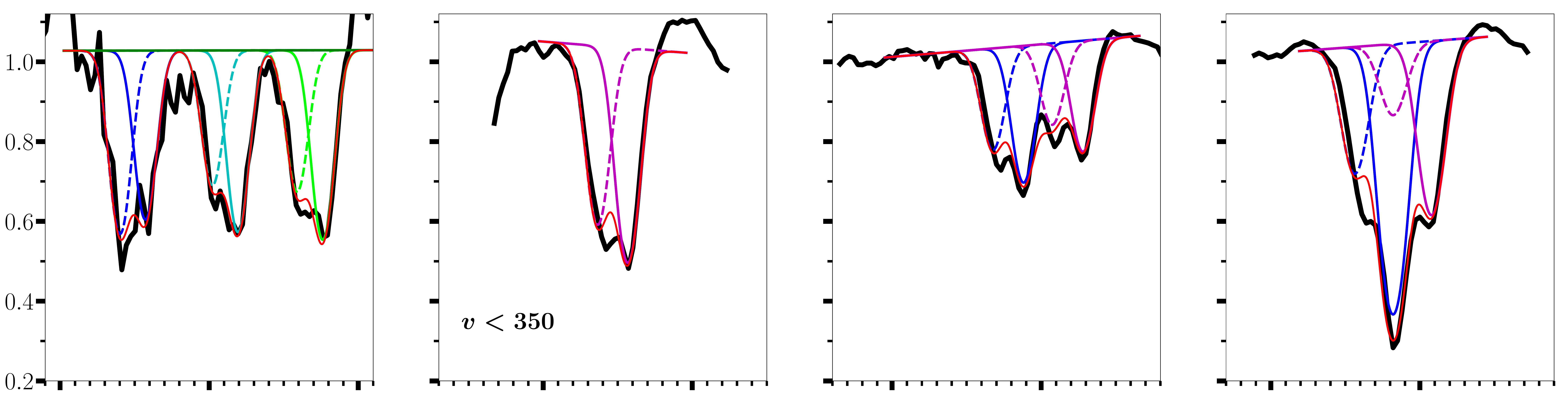}}\
  \raisebox{\dimexpr-.5\height-1em}{\includegraphics[scale=0.17]{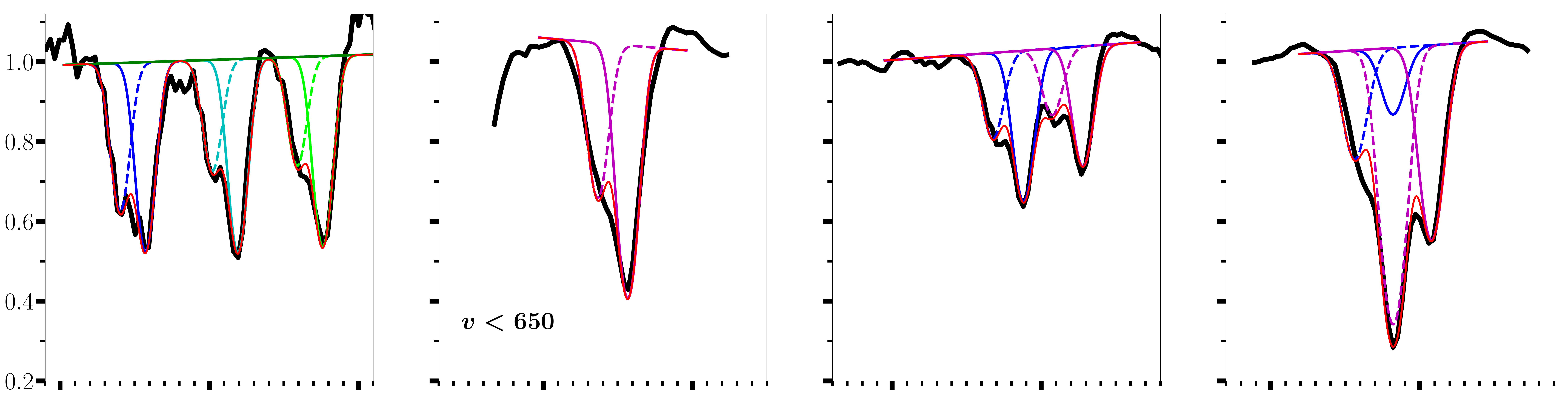}}\
  \raisebox{\dimexpr-.5\height-1em}{\includegraphics[scale=0.17]{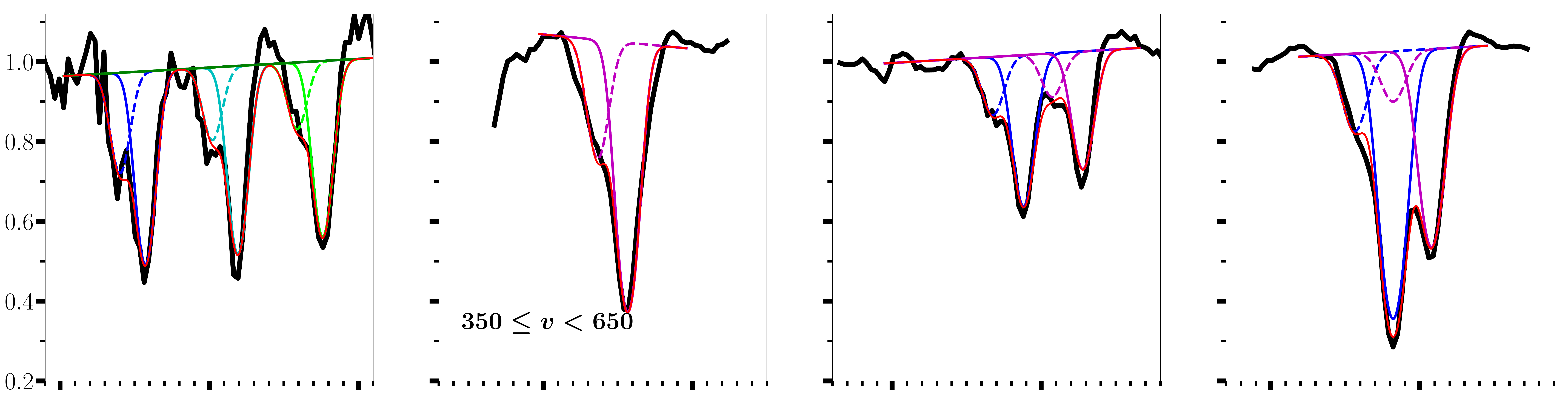}}\
  \raisebox{\dimexpr-.5\height-1em}{\includegraphics[scale=0.17]{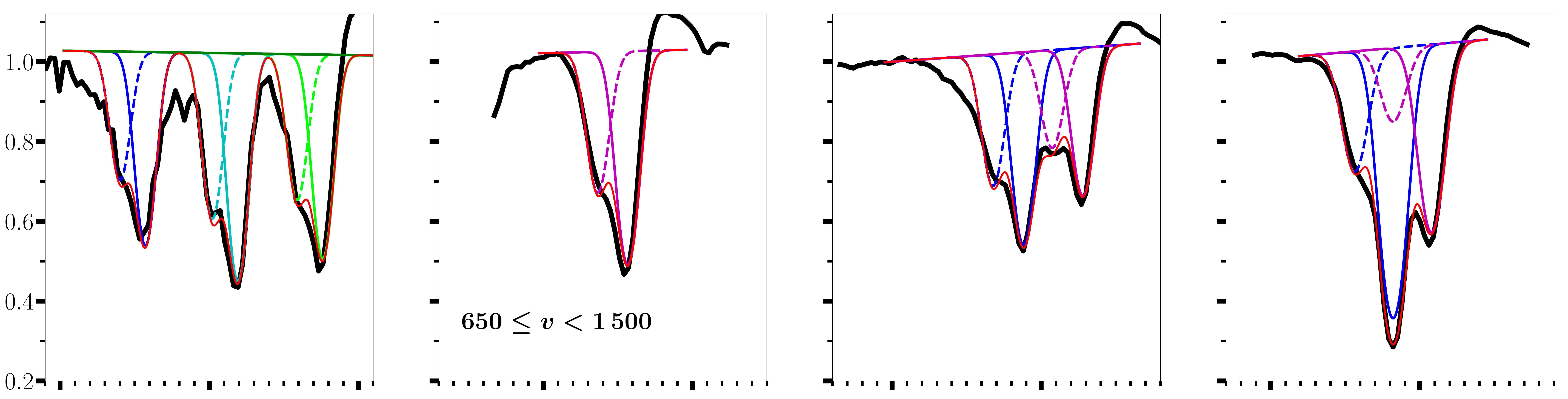}}\
  \raisebox{\dimexpr-.5\height-1em}{\includegraphics[scale=0.17]{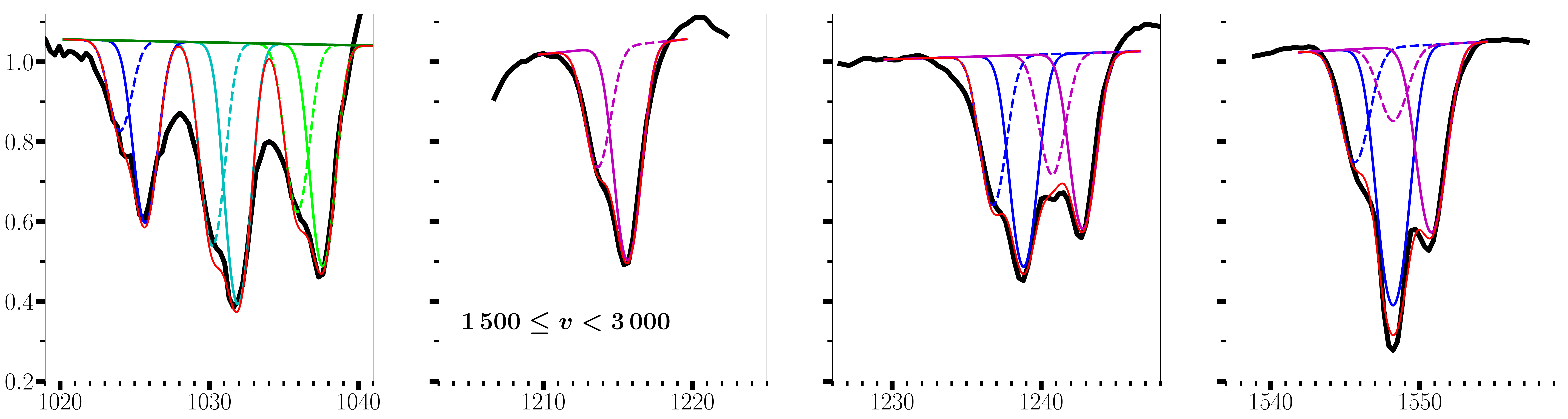}}\

  \vspace*{0.2cm}\small\textcolor{black}{Outflow rest-frame wavelength [\r{A}]}
  \caption{Fits to the strongest features for the composite outflow spectra in Table 2, as in Figure \ref{fig:fitlock}.}\label{fig:fitall}
\end{minipage}%
\end{figure}
\begin{figure}
\centering
\begin{minipage}{0.3cm}
\rotatebox{90}{\small\textcolor{black}{Transmission}}
\end{minipage}%
\begin{minipage}{\dimexpr\linewidth-2.50cm\relax}%
    \centering
  \raisebox{\dimexpr-.5\height-1em}{\includegraphics[scale=0.17]{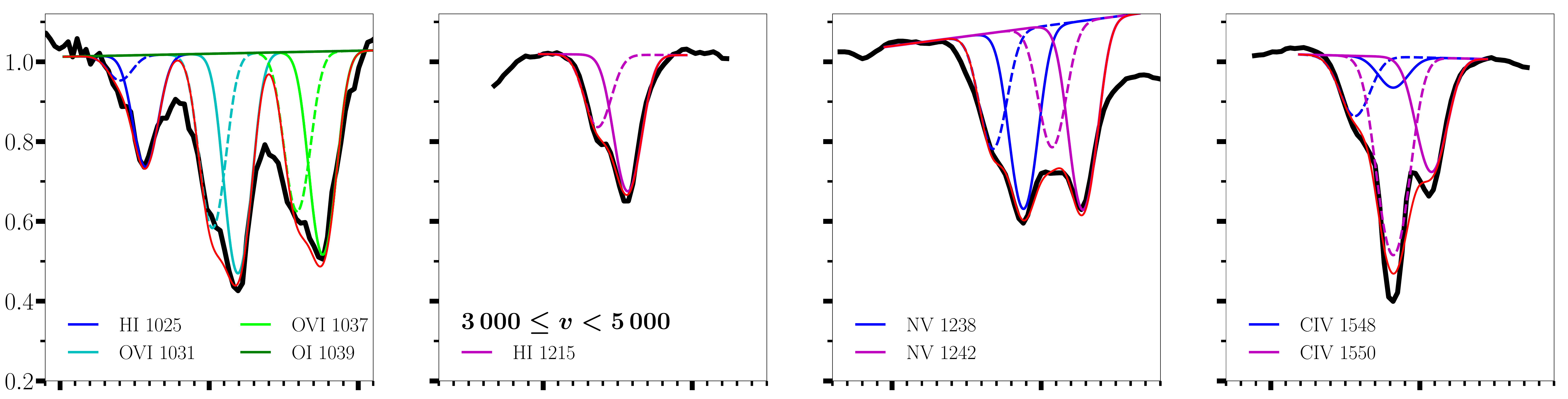}}\
  \raisebox{\dimexpr-.5\height-1em}{\includegraphics[scale=0.17]{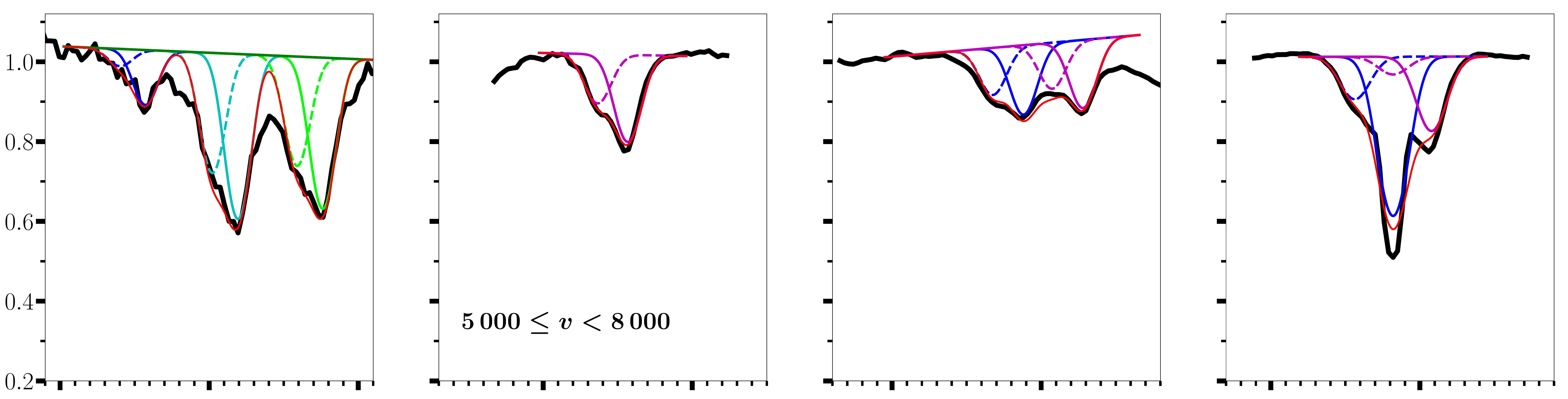}}\
  \raisebox{\dimexpr-.5\height-1em}{\includegraphics[scale=0.17]{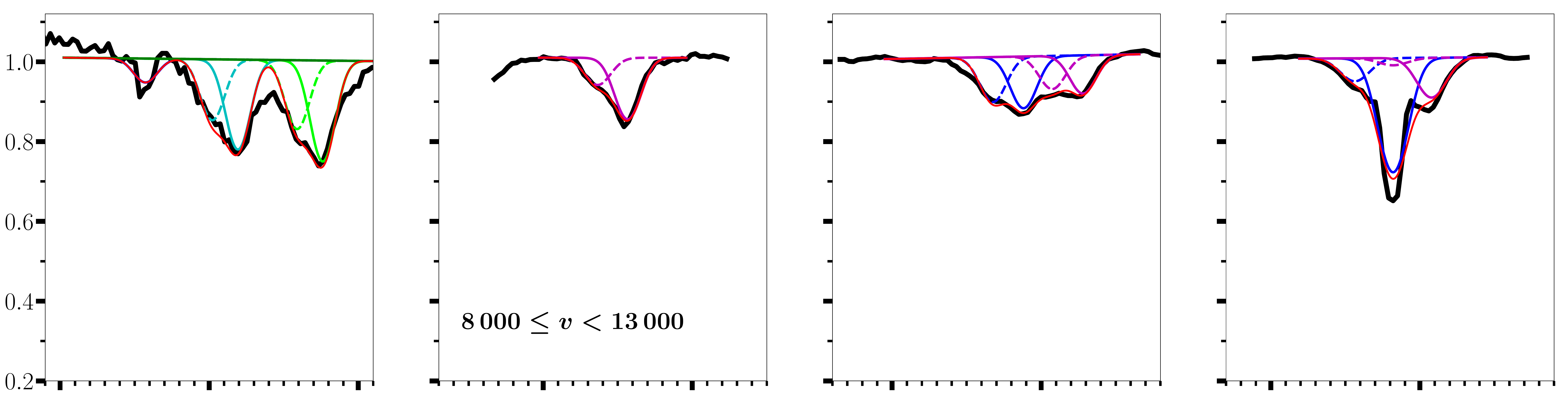}}\
  \raisebox{\dimexpr-.5\height-1em}{\includegraphics[scale=0.17]{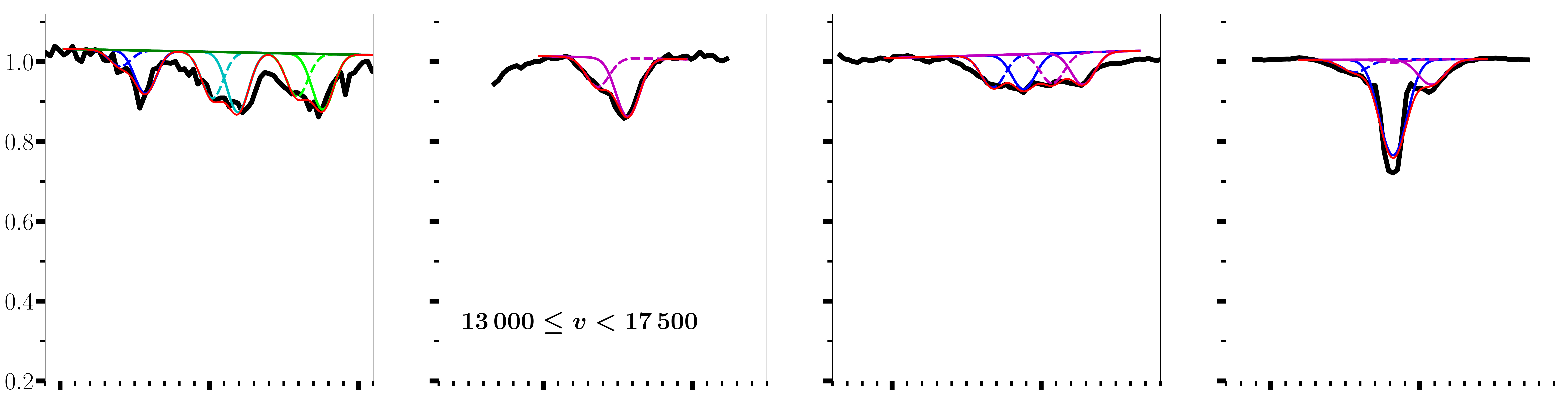}}\
  \raisebox{\dimexpr-.5\height-1em}{\includegraphics[scale=0.17]{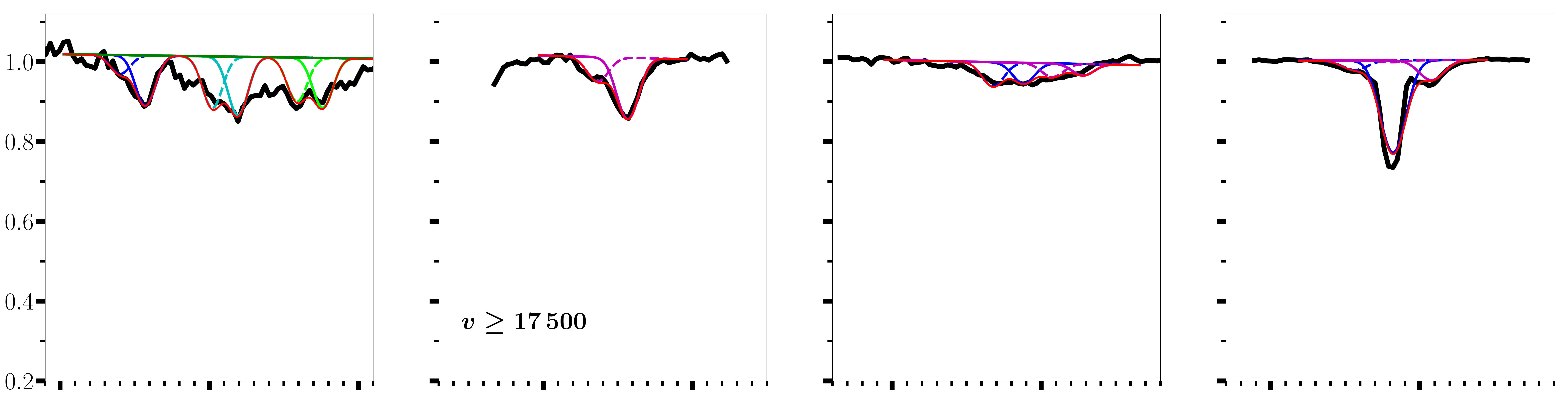}}\
  \raisebox{\dimexpr-.5\height-1em}{\includegraphics[scale=0.17]{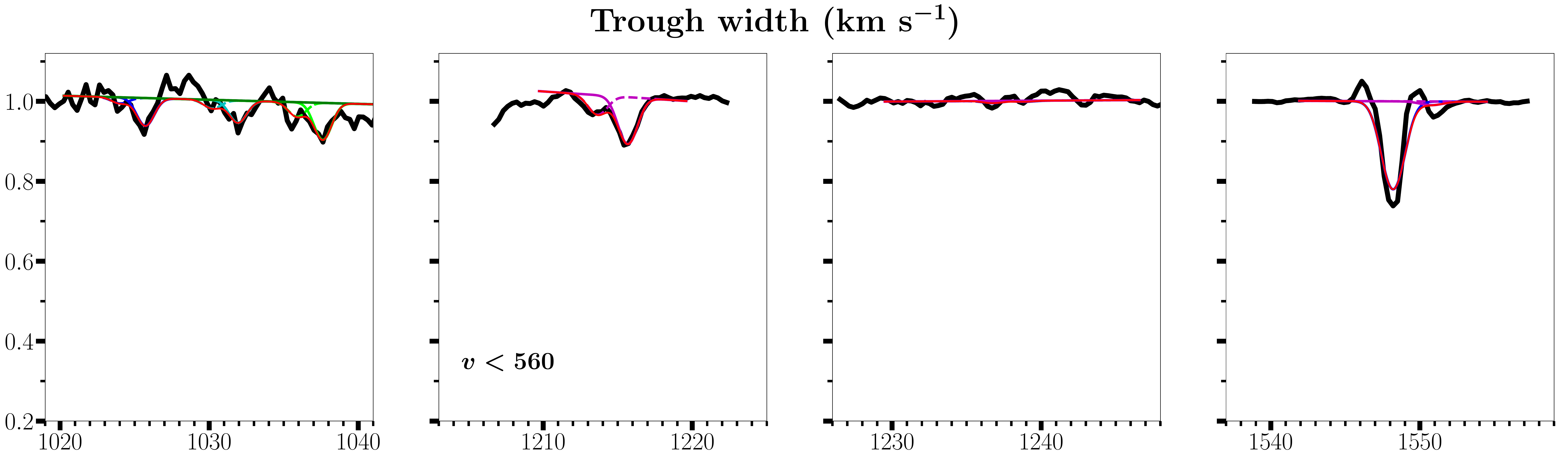}}\

  \vspace*{0.2cm}\small\textcolor{black}{Outflow rest-frame wavelength [\r{A}]}
  \caption{Continued}
\end{minipage}%
\end{figure}
\begin{figure}
\centering
\begin{minipage}{0.3cm}
\rotatebox{90}{\small\textcolor{black}{Transmission}}
\end{minipage}%
\begin{minipage}{\dimexpr\linewidth-2.50cm\relax}%
    \centering
  \raisebox{\dimexpr-.5\height-1em}{\includegraphics[scale=0.17]{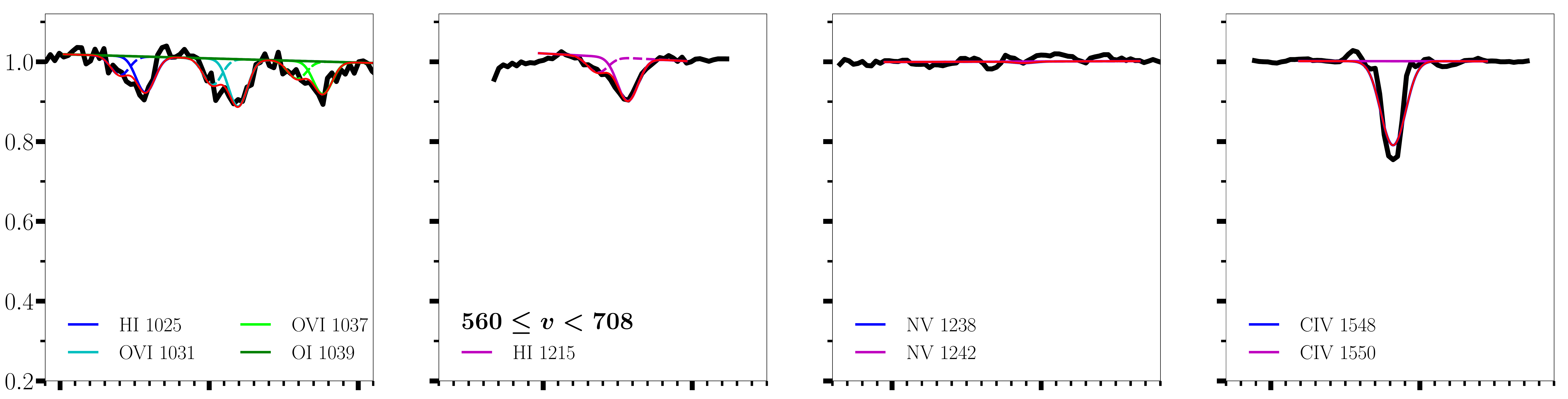}}\
  \raisebox{\dimexpr-.5\height-1em}{\includegraphics[scale=0.17]{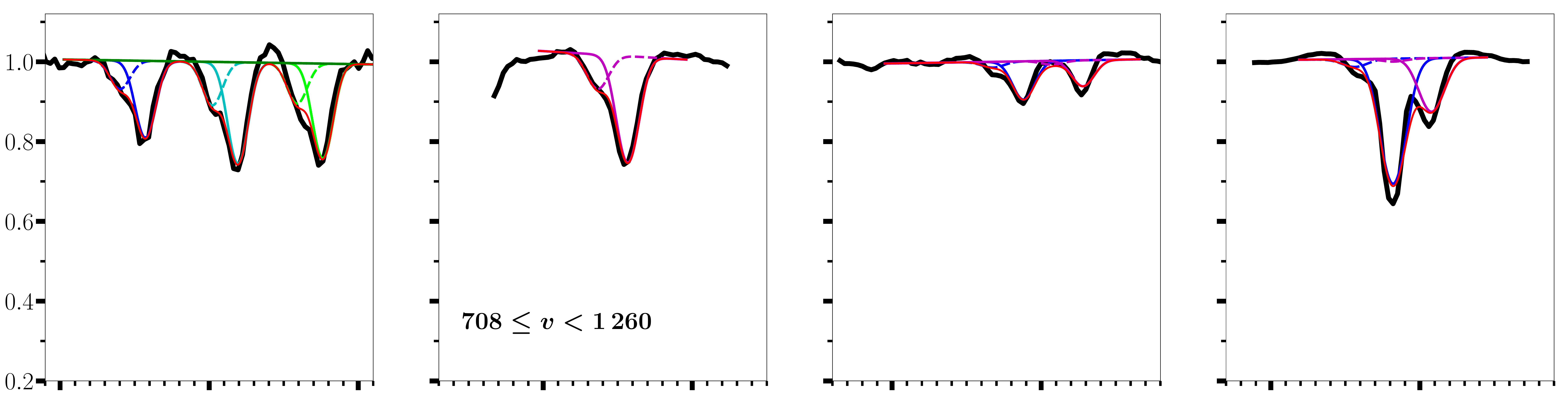}}\
  \raisebox{\dimexpr-.5\height-1em}{\includegraphics[scale=0.17]{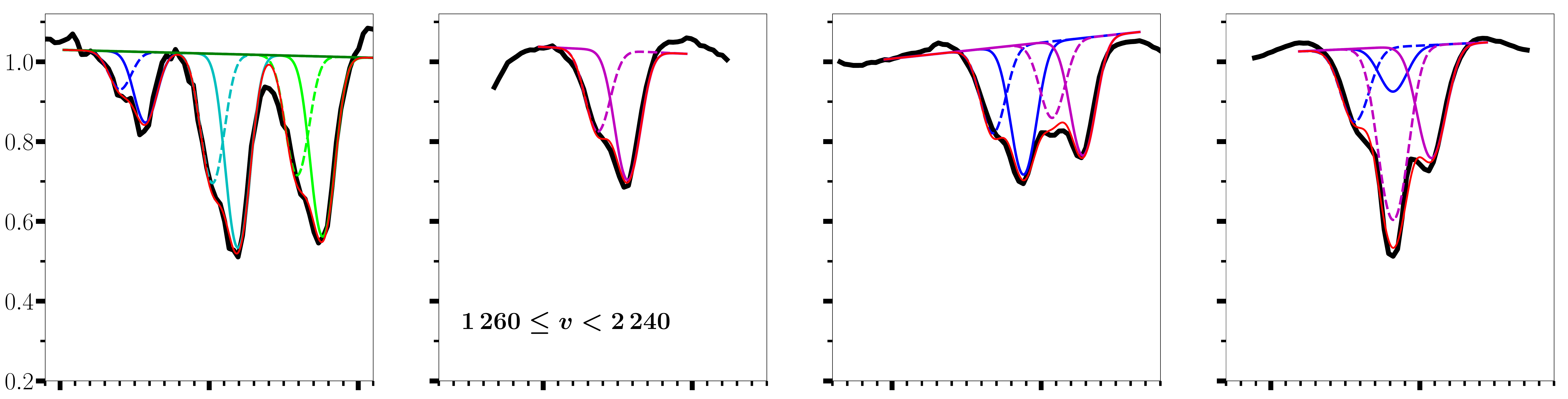}}\
  \raisebox{\dimexpr-.5\height-1em}{\includegraphics[scale=0.17]{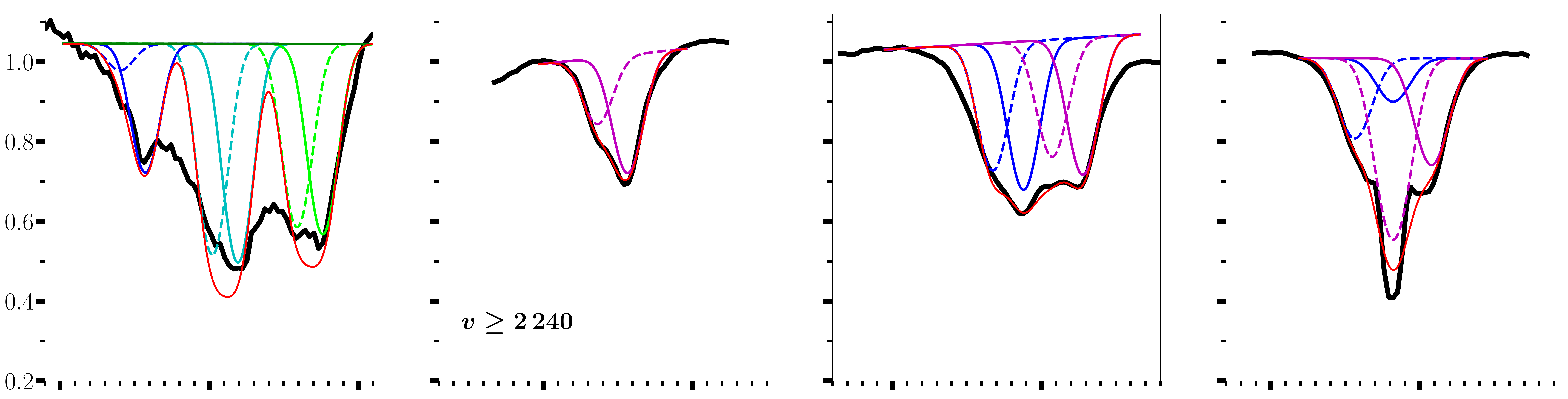}}\
  \raisebox{\dimexpr-.5\height-1em}{\includegraphics[scale=0.17]{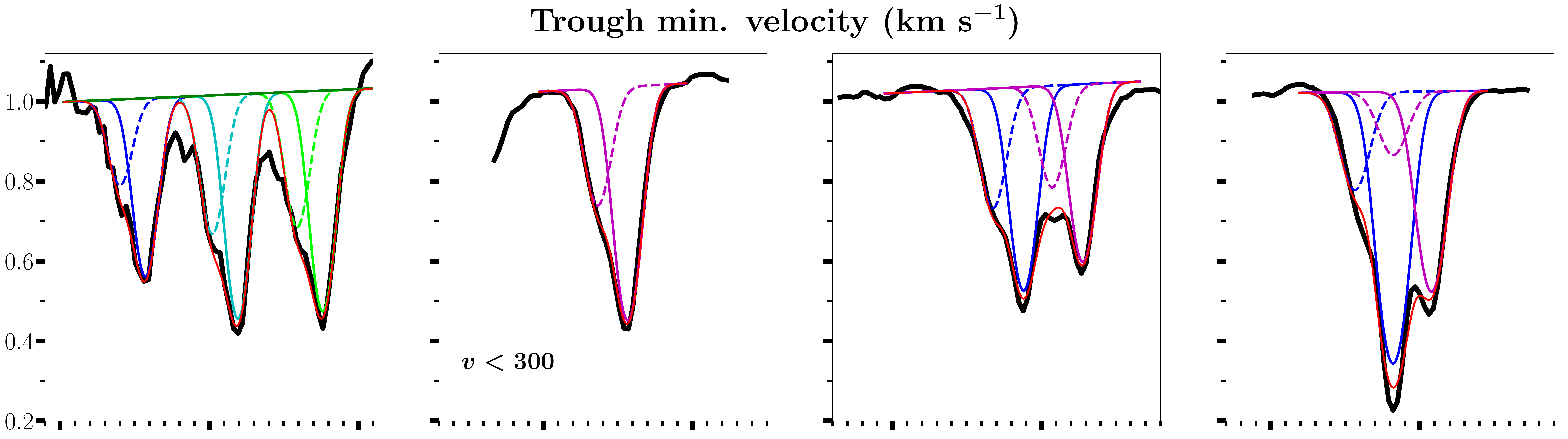}}\
  \raisebox{\dimexpr-.5\height-1em}{\includegraphics[scale=0.17]{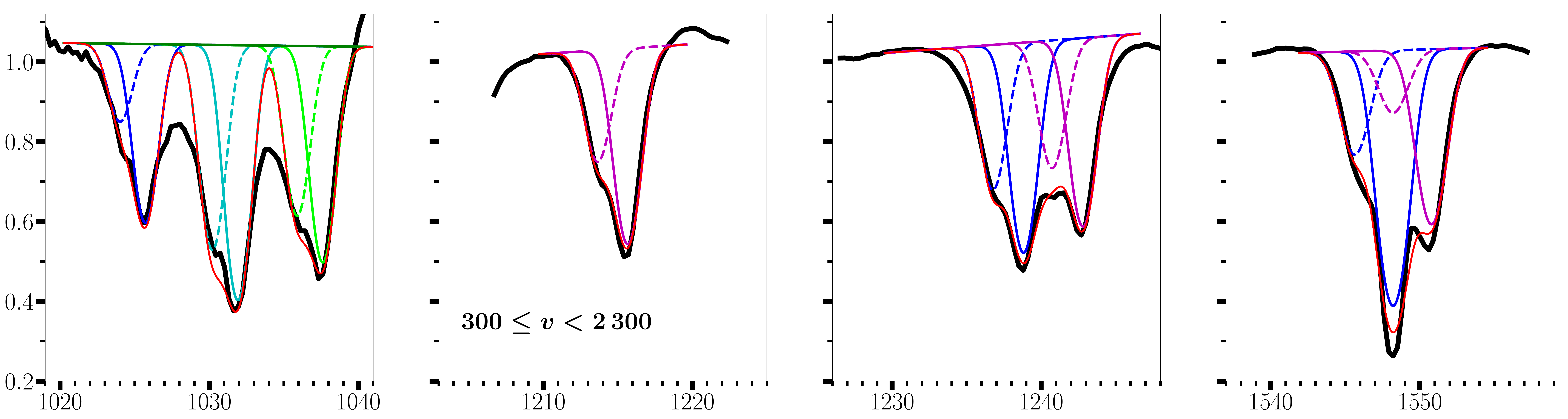}}\

  \vspace*{0.2cm}\small\textcolor{black}{Outflow rest-frame wavelength [\r{A}]}
  \caption{Continued}
\end{minipage}%
\end{figure}
\begin{figure}
\centering
\begin{minipage}{0.3cm}
\rotatebox{90}{\small\textcolor{black}{Transmission}}
\end{minipage}%
\begin{minipage}{\dimexpr\linewidth-2.50cm\relax}%
    \centering
  \raisebox{\dimexpr-.5\height-1em}{\includegraphics[scale=0.17]{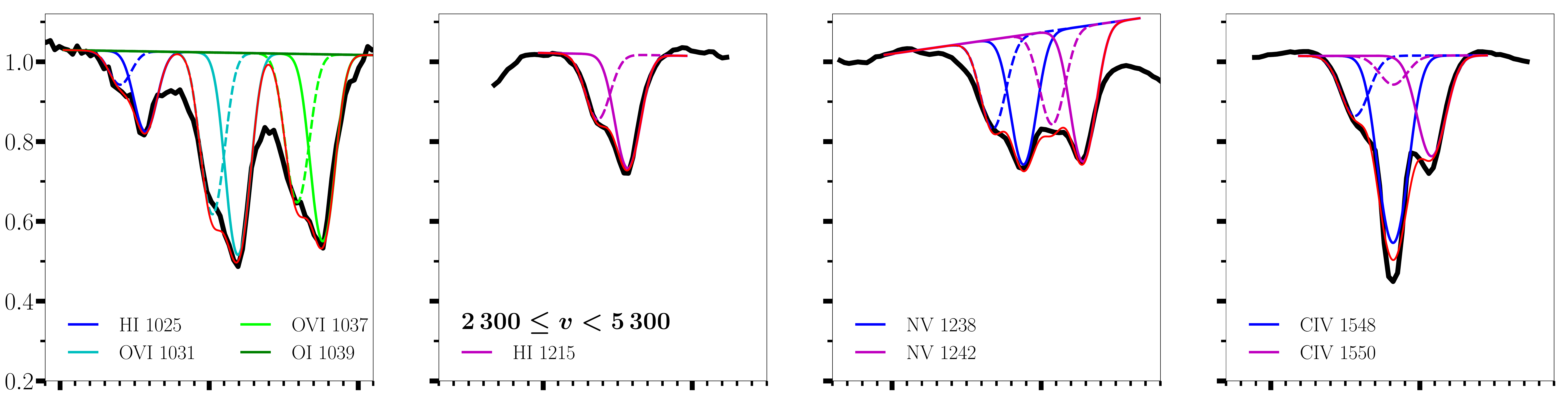}}\
  \raisebox{\dimexpr-.5\height-1em}{\includegraphics[scale=0.17]{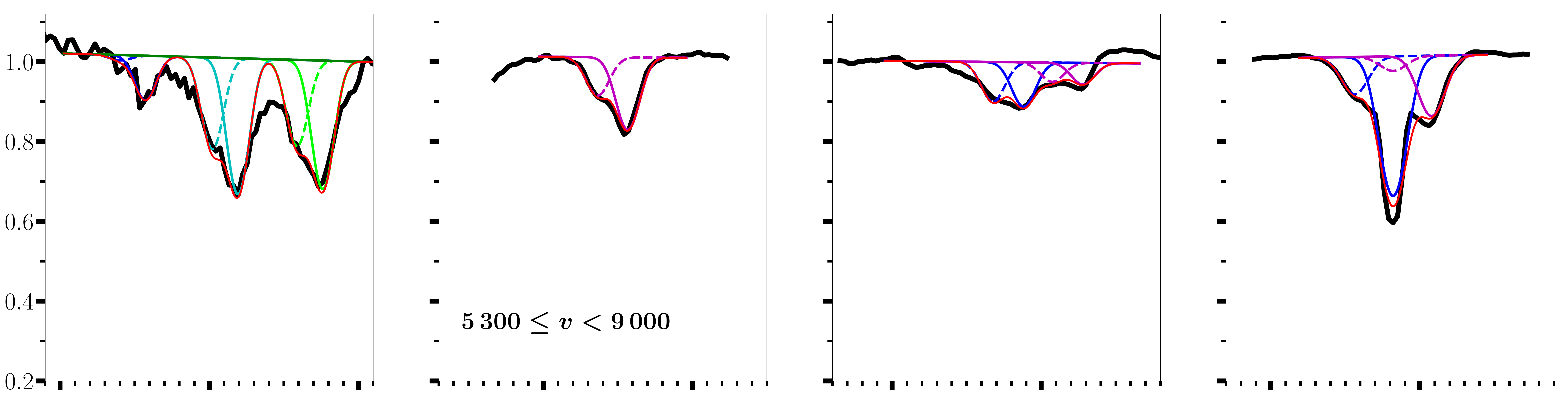}}\
  \raisebox{\dimexpr-.5\height-1em}{\includegraphics[scale=0.17]{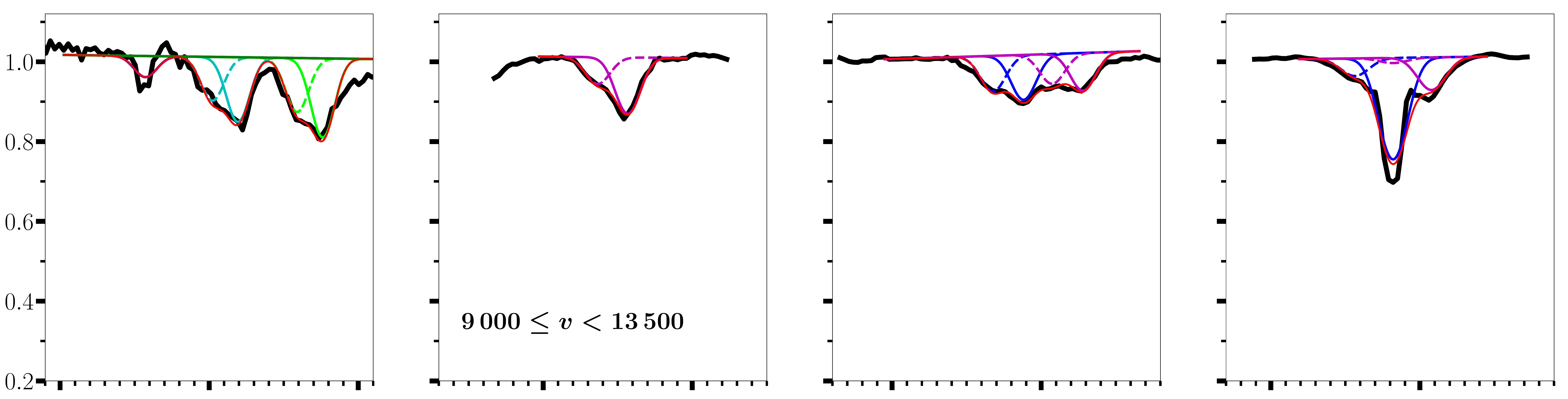}}\
  \raisebox{\dimexpr-.5\height-1em}{\includegraphics[scale=0.17]{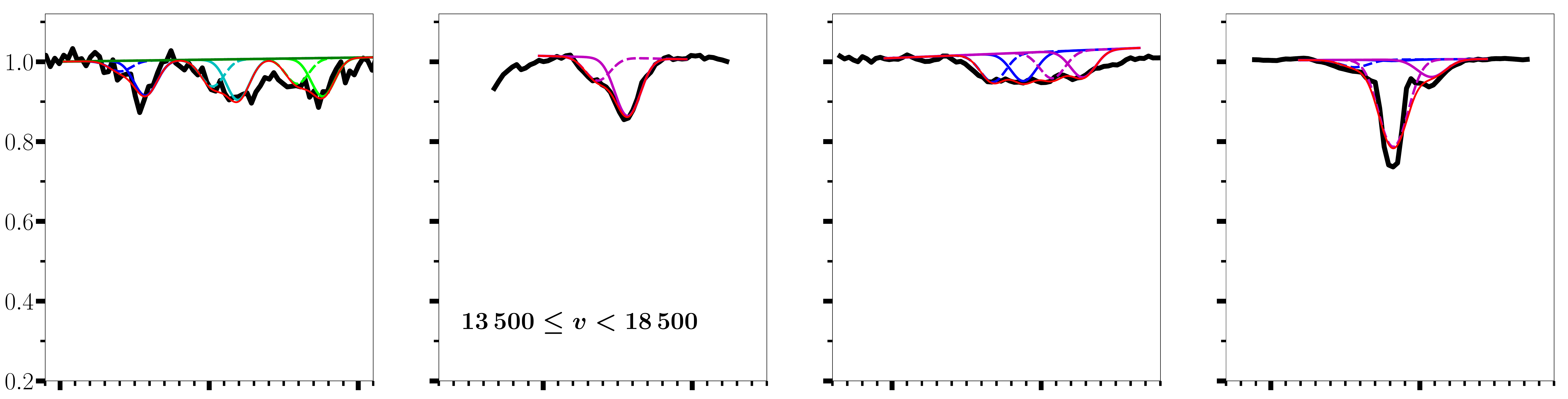}}\
  \raisebox{\dimexpr-.5\height-1em}{\includegraphics[scale=0.17]{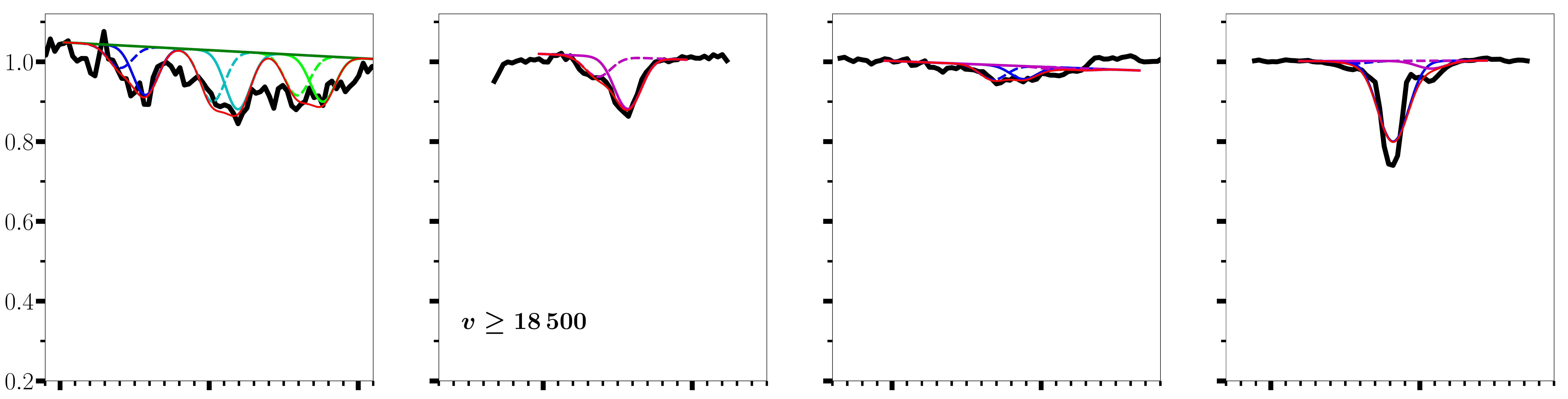}}\
  \raisebox{\dimexpr-.5\height-1em}{\includegraphics[scale=0.605]{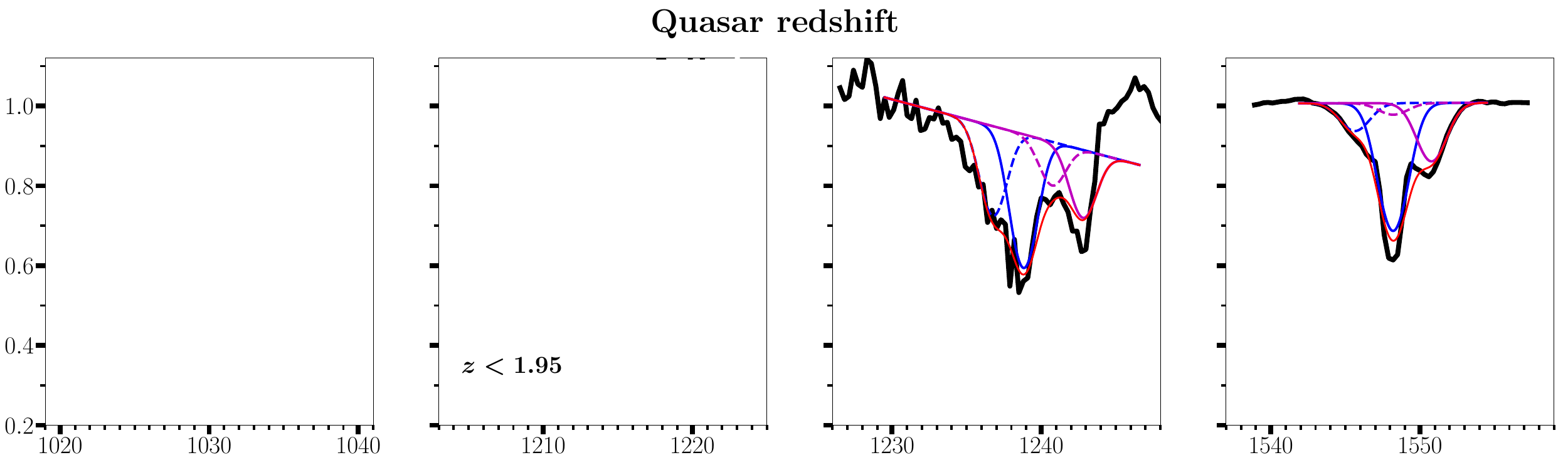}}\

  \vspace*{0.2cm}\small\textcolor{black}{Outflow rest-frame wavelength [\r{A}]}
  \caption{Continued}
\end{minipage}%
\end{figure}
\begin{figure}
\centering
\begin{minipage}{0.3cm}
\rotatebox{90}{\small\textcolor{black}{Transmission}}
\end{minipage}%
\begin{minipage}{\dimexpr\linewidth-2.50cm\relax}%
    \centering
  \raisebox{\dimexpr-.5\height-1em}{\includegraphics[scale=0.17]{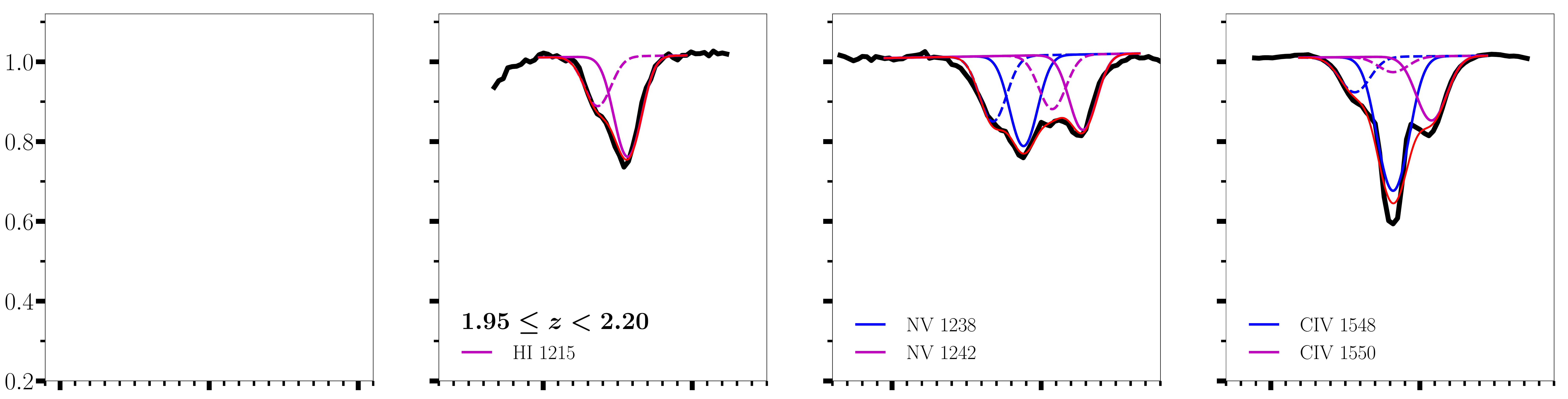}}\
  \raisebox{\dimexpr-.5\height-1em}{\includegraphics[scale=0.17]{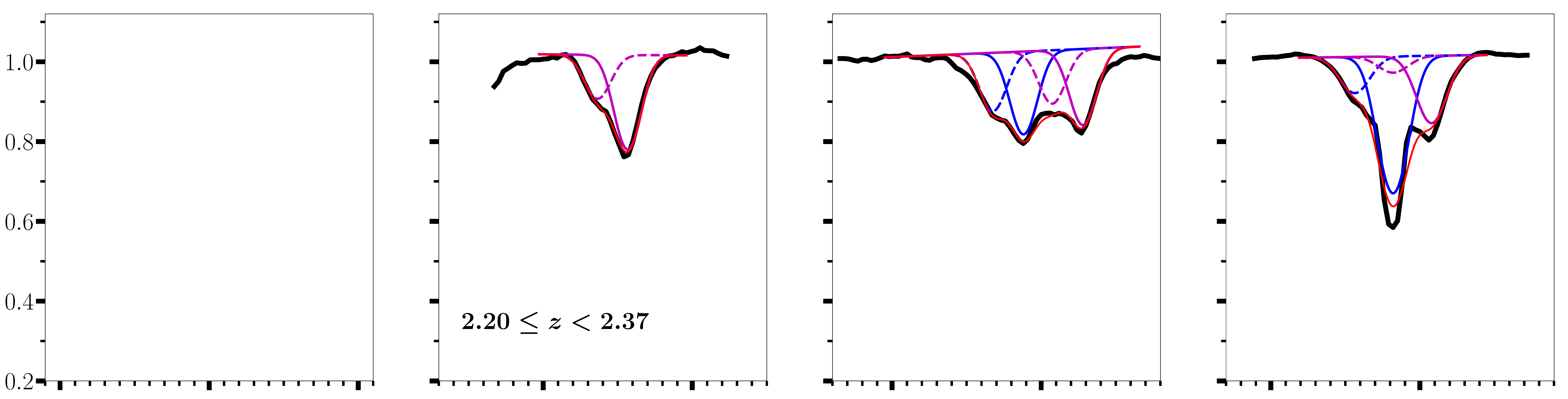}}\
  \raisebox{\dimexpr-.5\height-1em}{\includegraphics[scale=0.17]{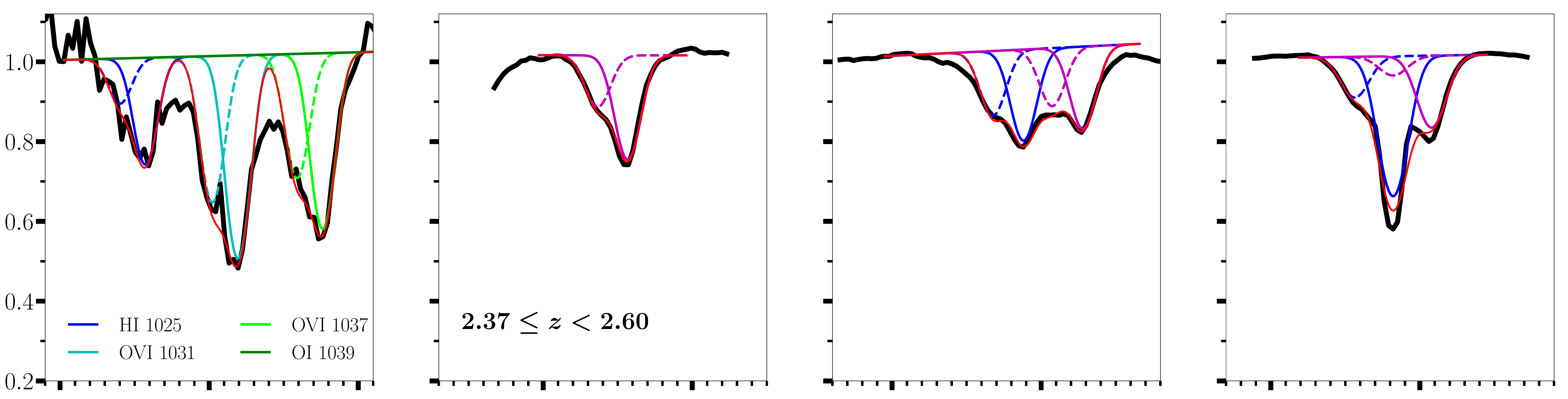}}\
  \raisebox{\dimexpr-.5\height-1em}{\includegraphics[scale=0.17]{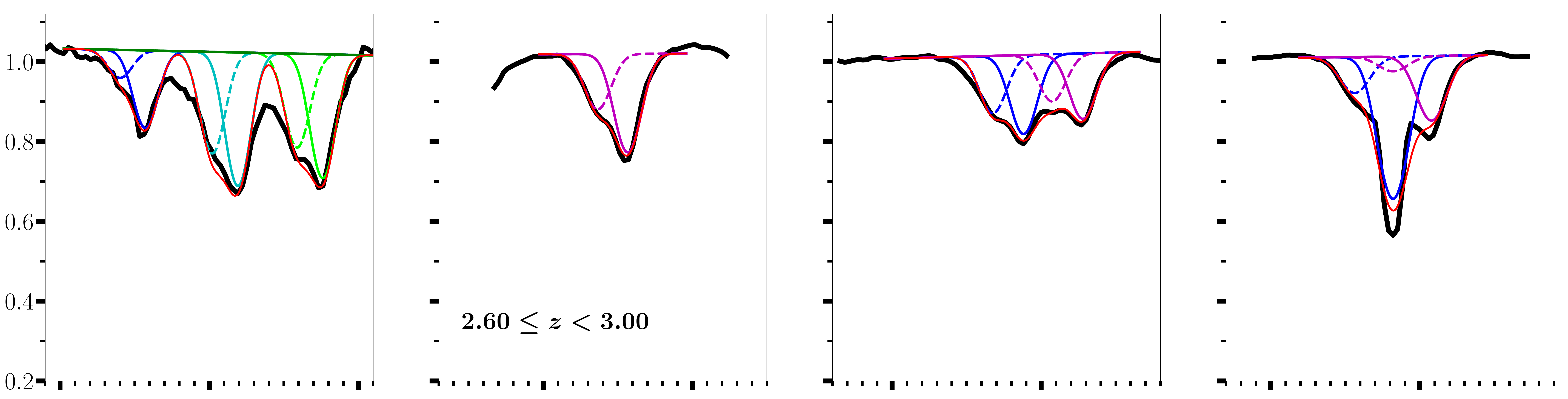}}\
  \raisebox{\dimexpr-.5\height-1em}{\includegraphics[scale=0.17]{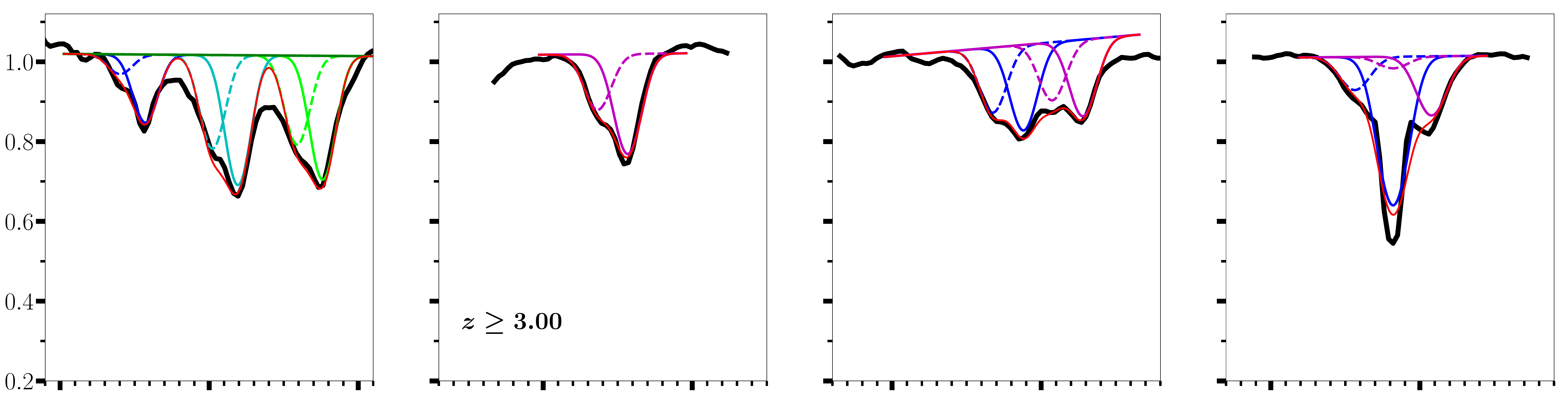}}\
  \raisebox{\dimexpr-.5\height-1em}{\includegraphics[scale=0.17]{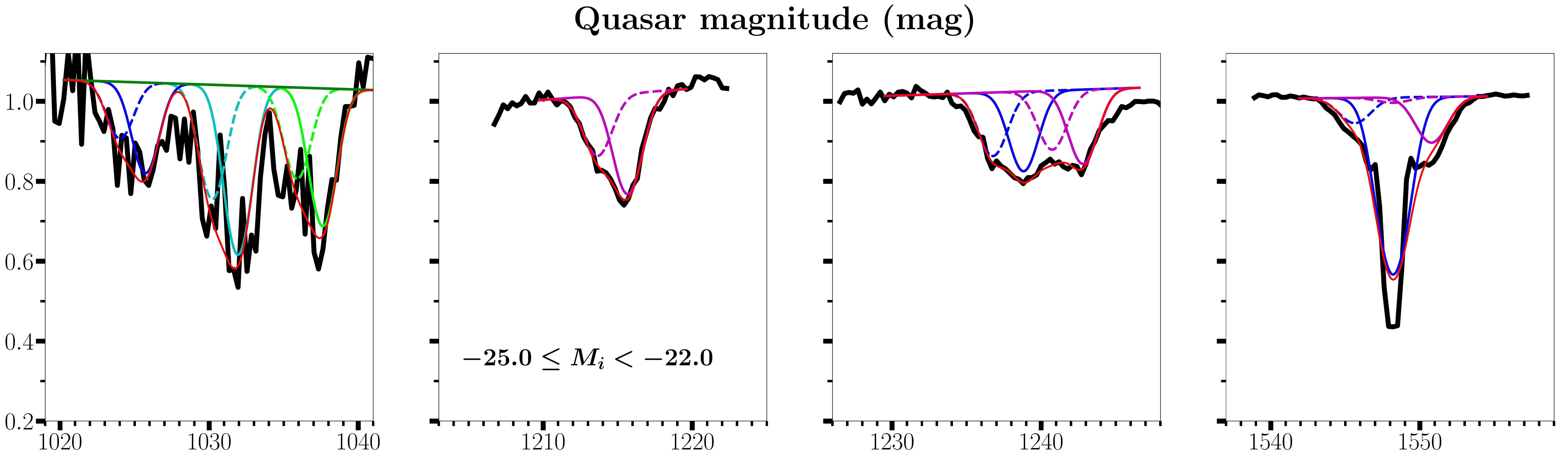}}\

  \vspace*{0.2cm}\small\textcolor{black}{Outflow rest-frame wavelength [\r{A}]}
  \caption{Continued}
\end{minipage}%
\end{figure}
\begin{figure}
\centering
\begin{minipage}{0.3cm}
\rotatebox{90}{\small\textcolor{black}{Transmission}}
\end{minipage}%
\begin{minipage}{\dimexpr\linewidth-2.50cm\relax}%
    \centering
  \raisebox{\dimexpr-.5\height-1em}{\includegraphics[scale=0.17]{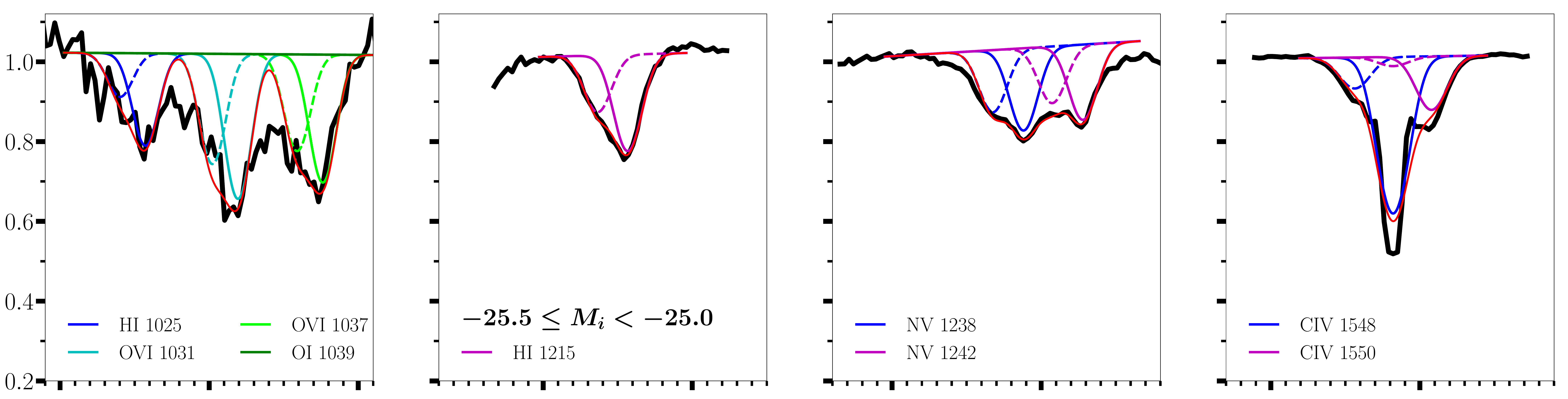}}\
  \raisebox{\dimexpr-.5\height-1em}{\includegraphics[scale=0.17]{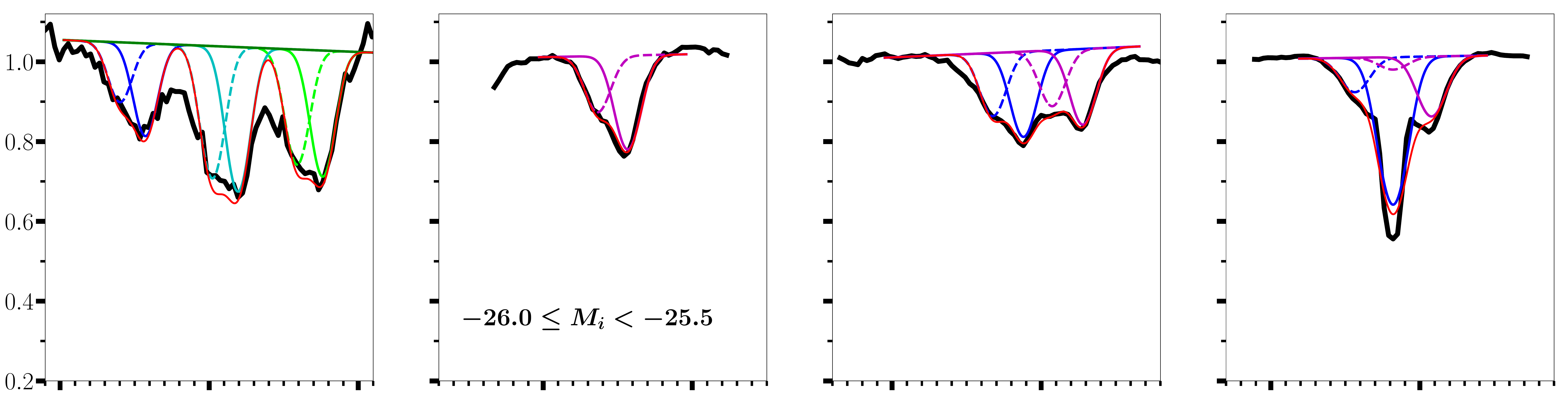}}\
  \raisebox{\dimexpr-.5\height-1em}{\includegraphics[scale=0.17]{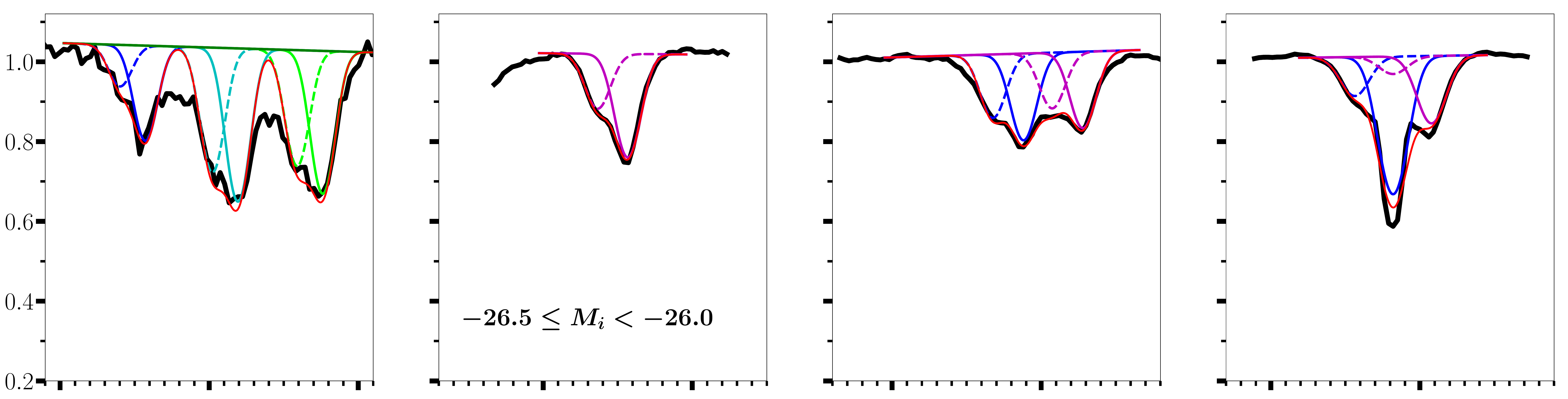}}\
  \raisebox{\dimexpr-.5\height-1em}{\includegraphics[scale=0.17]{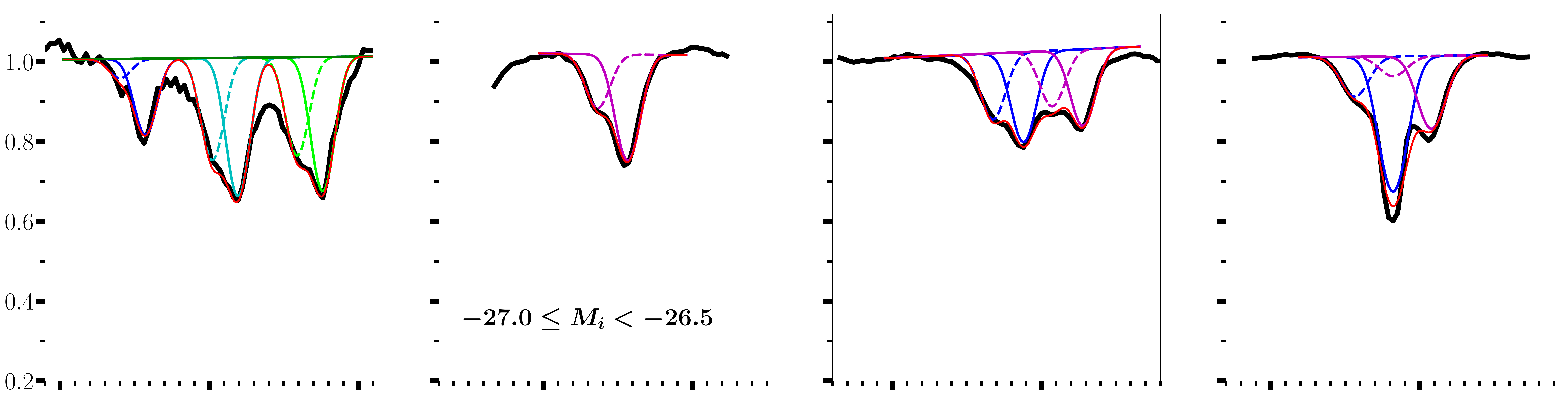}}\
  \raisebox{\dimexpr-.5\height-1em}{\includegraphics[scale=0.17]{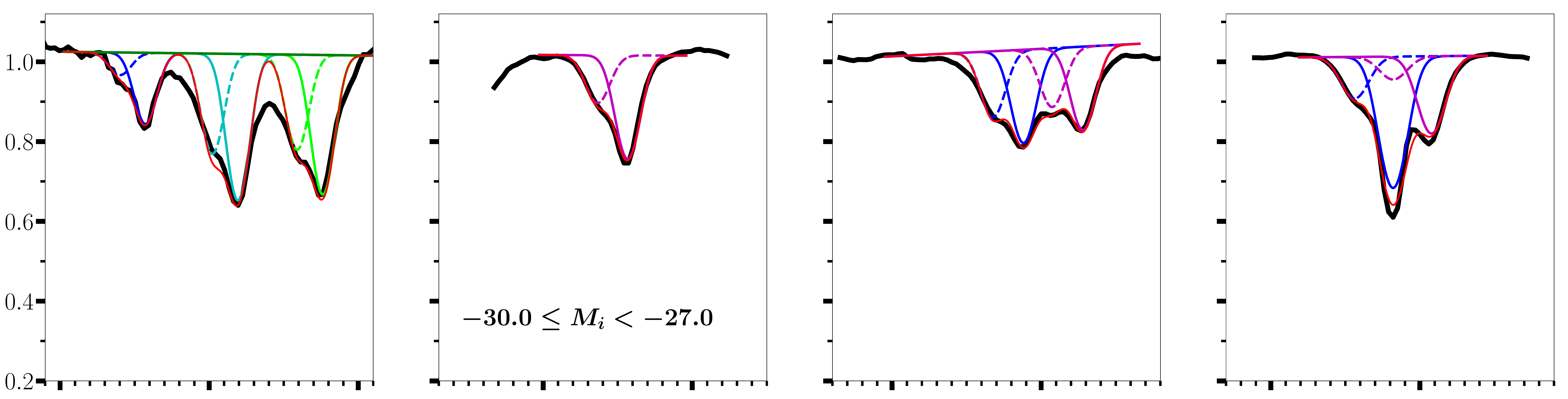}}\
  \raisebox{\dimexpr-.5\height-1em}{\includegraphics[scale=0.17]{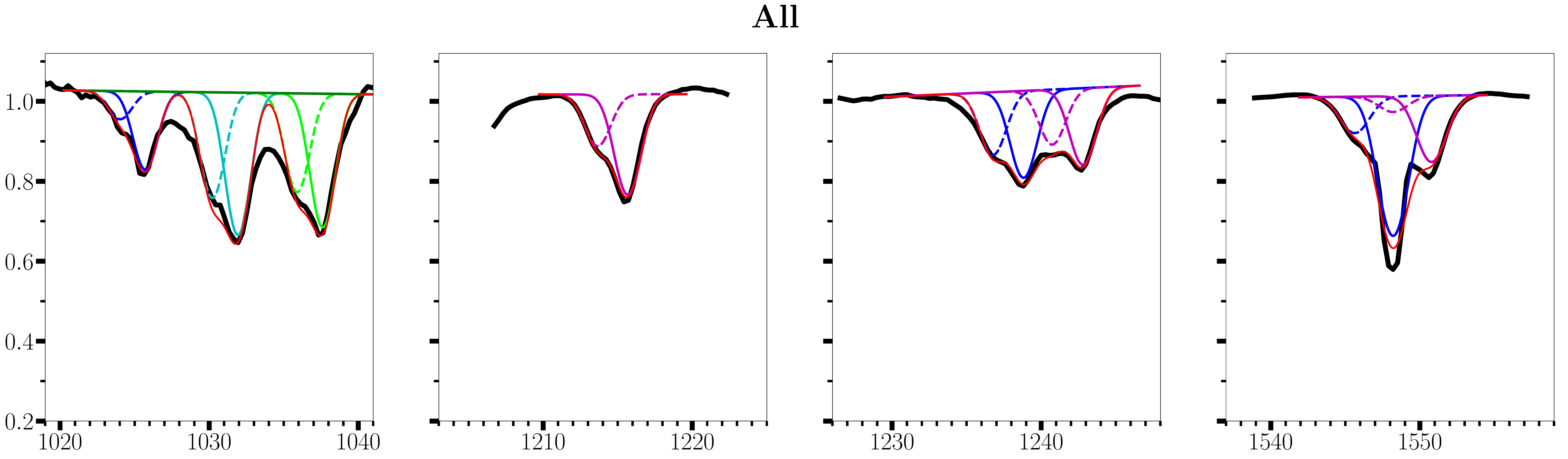}}\

  \vspace*{0.2cm}\small\textcolor{black}{Outflow rest-frame wavelength [\r{A}]}
  \caption{Continued}
\end{minipage}%
\end{figure}

\end{document}